\documentclass[10pt,journal,compsoc]{IEEEtran}
\usepackage{amssymb}
\usepackage{graphics,color}
\usepackage{graphicx}
\usepackage{amssymb,amsmath,amsfonts,amssymb}
\usepackage{amsmath,amssymb,mathrsfs,bm,url,times}
\usepackage{subfigure}
\usepackage{extarrows}
	\usepackage{graphicx,algorithm,algorithmic}
	\usepackage{graphicx,url}
\usepackage{cases}
	\newtheorem{The}{Theorem}
	\newtheorem{Pro}[The]{Proposition}

	\newtheorem{Def}[The]{Definition}
	\newtheorem{Exa}[The]{Example}

	\numberwithin{equation}{section} \numberwithin{The}{section}

	\newcommand{\p}{{\rm P}}
	
	\newcommand{\ignore}[1]{}

	\def\proof{\trivlist \item[\hskip \labelsep{\bf Proof\ }]}

		\def\qed{\hfill \vrule height 6pt width 6pt depth 0pt}

\ifCLASSOPTIONcompsoc
  \usepackage[nocompress]{cite}
\else
  \usepackage{cite}
\fi
\ifCLASSINFOpdf
\else
\fi
\begin{document}
%
\title{Joint Cyber Risk   Assessment  of Network Systems with Heterogeneous Components}
%
%
%
%

\author{Gaofeng~Da, Maochao~Xu,  Jingshi~ Zhang, and~Peng~Zhao
\IEEEcompsocitemizethanks{\IEEEcompsocthanksitem Gaofeng Da and Jingshi Zhang are with the College of Economics and Management,
 Nanjing University of Aeronautics and Astronautics, China. Email:  dagfvc@gmail.com. \protect \\
\IEEEcompsocthanksitem Maochao Xu is with the Department of Mathematics,
    Illinois State University, USA. E-mail: mxu2@ilstu.edu. \protect \\
\IEEEcompsocthanksitem Peng Zhao is with School of Mathematics and Statistics,
      Jiangsu Normal University,  China.
      Corresponding author. E-mail: zhaop@jsnu.edu.cn.\protect\\
    }
}

\IEEEtitleabstractindextext{%
\begin{abstract}

 Cyber risks are the most common risks encountered by a modern network system. However, it is significantly difficult to assess the joint cyber risk owing to the network topology, risk propagation, and heterogeneities of components. In this paper, we propose a novel backward elimination approach for computing the joint cyber risk encountered by different types of components in a network system; moreover, explicit formulas are also presented. Certain specific network topologies including complete, star, and complete bi-partite topologies are studied. The effects of propagation depth and compromise probabilities on the joint cyber risk are analyzed using stochastic comparisons. The variances and correlations of cyber risks are examined by a simulation experiment. It was discovered that both variances and correlations change rapidly when the propagation depth increases from its initial value. Further, numerical examples are also presented.

\end{abstract}

}

\maketitle

\IEEEdisplaynontitleabstractindextext

%
\IEEEpeerreviewmaketitle

\IEEEraisesectionheading{\section{Introduction}\label{sec1}}

%
%
%
%
\IEEEPARstart{N}{etwork
} systems have become an indispensable component in modern society. They are important owing to the rapid evolution of cyber infrastructure, which typically consists of different processors, eStorage devices, sensors, and computers. For example, a key component for operating the cyber infrastructure is the supervisory control and data acquisition (SCADA) system, which performs monitoring, analyzing, and controlling tasks by using computers and networked data communications \cite{igure2006security}. Another example is the rapid development of the internet of things, which relies on the cyber network system to connect and exchange data. The network system consists of sensors, computers, e-devices, and other objects to gather and share the information. While the network systems are essential and beneficial to society, they encounter significant cyber risks \cite{ten2008vulnerability, suo2012security}. According to the Repository of Industrial Security Incidents ({\url{http://www.risidata.com/}), 242 security incidents related to critical infrastructure and industrial control systems have occurred from 1982 to 2014 \cite{ogie2017cyber}. Therefore, there is an urgent demand for the methodologies to assess the cyber risks of network systems, which is challenging owing to the dependence among risks and heterogeneous nature of the network system (i.e., different types of components).
	

	In the literature, there are extensive studies on the risk assessment of network systems \cite{zou2007modeling, xu2011stochastic, cherdantseva2016review,yang2018risk}. For instance, Cherdantseva et al. \cite{cherdantseva2016review} reviewed twenty-four risk assessment methods pertaining to a SCADA system, and various approaches were discussed. A comprehensive review of the studies on the epidemic spreading over complex networks was presented in \cite{pastor2015epidemic}. Although there are extensive models for studying the cyber risks of networked systems, the dependence among the risks were not thoroughly understood despite its natural existence among the cyber risks of network systems. For example, for a smart grid with a SCADA system, the risks encountered by the computers and phasor measurement units (PMUs) are highly correlated as the attacks can propagate through the communication network. The challenge existing in the dependence study is primarily caused by the technique barrier as the network system typically involves different types of components. There are only a few works that analyze joint/dependent risks. Xu and Xu \cite{xu2012extended} proposed a stochastic model to assess the risks of a network system, where the stochastic renewal process was utilized to handle the joint risk. Xu et al. \cite{xu2015cyber} introduced the tool of copulas to accommodate the dependencies among the cyber risks over a complex network, where the copula is a statistical tool that can handle the nonlinear dependence and is widely used in several different areas \cite{joe2014dependence}. Qu and Wang \cite{qu2017sis} developed a correlated heterogeneous susceptible-infectious-susceptible (SIS) model over a network, where the infection rates were assumed to be correlated to nodal degrees. Laszka et al. \cite{laszka2018assessment} studied a correlated risk model over a network system from the cyber insurance perspective, where the probability distribution of the total number of compromised nodes within a network was provided. To the best of our knowledge, none of the existing studies have discussed the joint risk over a network system with heterogeneous nodes, which is the main purpose of this work.

To further understand the motivations for our study, consider the synchrophasor data system of a smart grid, as illustrated in Figure \ref{fig:pmu}, which has three different types of components \cite{phadke2008synchronized}: (i) a PMU monitors the working status of a power grid by recording the measurement data including voltage, current, frequency, and phase angle; (ii) a phasor data concentrator (PDC) coverts the phasor data obtained from multiple PMUs, which is outputted to the data stream; further, the data can be transmitted to other PDCs or to workstations; (iii) a workstation, where decisions are made and operations performed based on the data from PDCs.
	\begin{figure}[!htbp]
	\centering
	\includegraphics[width=.48\textwidth]{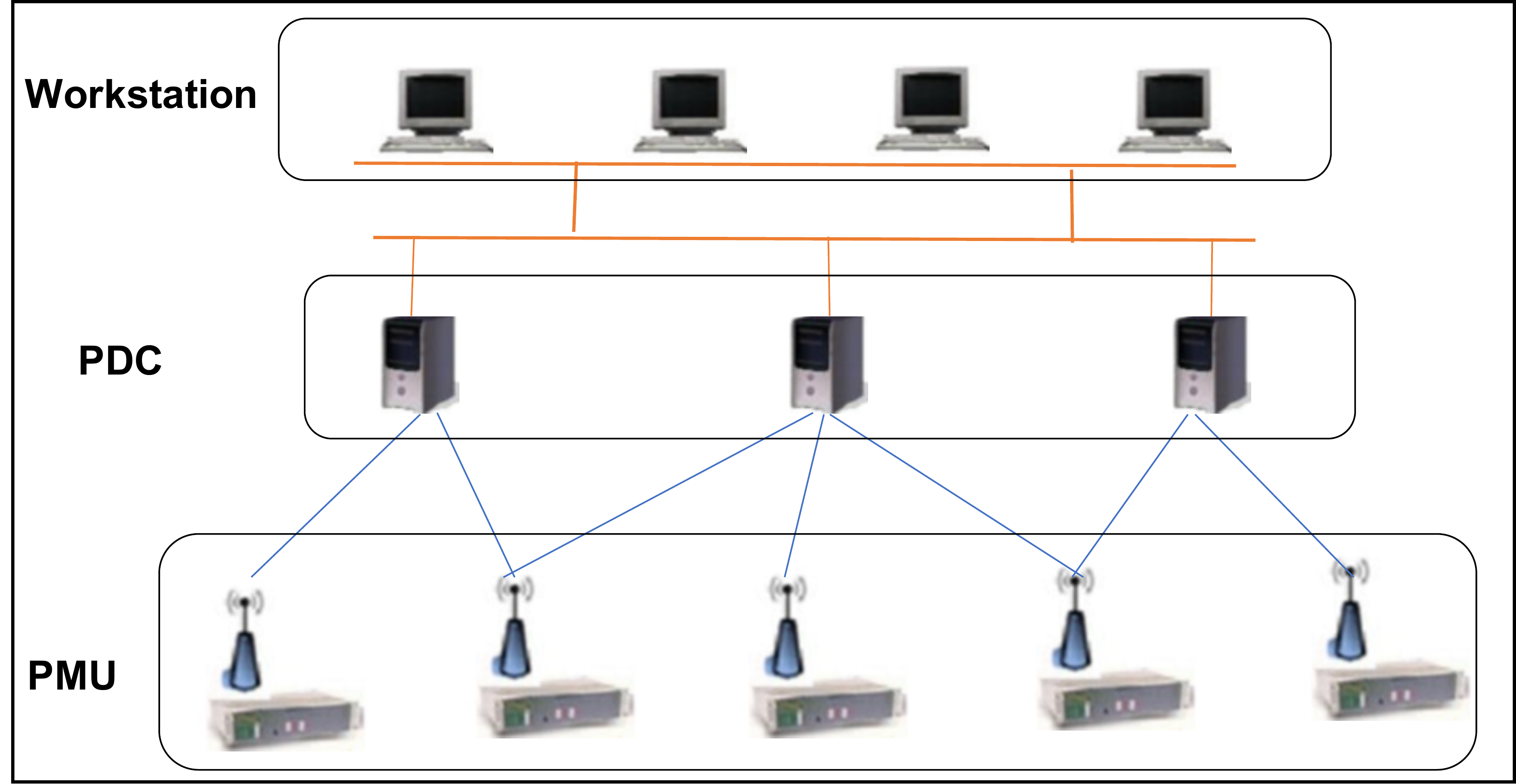}
	 \caption{Illustration of synchrophasor control system \label{fig:pmu}}
	\end{figure}
	 As those components are networked and the attacks can propagate over the network, it is particuarly important to study the joint risk of the different components. Specifically, the system defender is required to know the information on the numbers of PMUs, PDCs, and computers in the workstation that are compromised to assess the risk of the entire system. For example, the risk level of the {smart grid in Figure \ref{fig:pmu}} can be scored from zero to five based on the joint risk levels listed in Table \ref{table:risk-score}. The risk score is the highest (five) if any of the computers in the workstation is compromised, all the PMUs are compromised, or four PMUs and at least one PDC is compromised. To develop a similar risk scoring system, the joint distribution of risks is required, which is also the motivation for the current work. Such a scoring system can be used by an insurance company for the purpose of pricing.
	\begin{table}[hbtp!]
	\caption{ Sample risk scoring approach for the synchrophasor data system, where $X_1$ represents the number of compromised PMUs, $X_2$ represents the number of compromised PDCs, and $X_3$ denotes the number of compromised computers in the workstation\label{table:risk-score}}
		\centering
		\begin{tabular}{cccc}
	\hline
	 	Risk score & $(X_1,X_2,X_3)$   \\
			\hline
		   0&   $(0,0,0)$ \\
		     1&   $(1,0,0)$ \\
	         2&   $(2,0,0)$ \\
	      3&   $(3,0,0), (0,1,0)$, {$(0,2,0)$}\\
	    4&   {$(1,\ge 1,0), (2,\ge 1,0), (3,\ge 1,0),(4,0,0)$} \\
	    5&   $(0,0,\ge1),  (4,\ge 1,0),(5,0,0)$ \\	\hline
		
		\end{tabular}
	\label{tabb}
	\end{table}
	
{ In this paper, we present an analysis of the joint cyber risk  in a  network system with heterogenous components. The main contributions are summarized as follows.
\begin{itemize}
  \item The explicit formulas of joint cyber risks are provided for   network systems with heterogenous components. Those explicit  solutions are  particularly important for assessing the joint risk of critical cyber components in a small size network system. A novel {\em backward elimination} approach for computing the joint  risk of a network system is developed for this purpose.  For a large-scale network with heterogenous components, the proposed backward elimination approach can be used to effectively simulate the joint risks.

  \item A new cyber index (namely, propagation depth $L$) is introduced. This new index  $L$ allows us to describe
 the power of risk propagation or the defensive power of a network. Specifically, the network defender can use this index to determine the propagation power of a new risk (e.g., a new malware). If $L$ is large, the new risk is considered to have more power for propagating over the network. This index can also be used for assessing the defense level of a company network. If a  less contagious malware  can propagate more hops over a company network, the defense level of the network can be marked as low.  This also helps the insurance company to determine the risk score of the company who wants to purchase the cyber insurance policy.

 \item We rigorously prove the effects of propagation parameters on the cyber risk of a network with heterogenous components via stochastic comparisons. The simulation study is also presented to confirm the theoretical results and provide new insights.
\end{itemize}

 }

 The rest of paper is organized as follows. In Section \ref{sec: propagation}, we introduce the $L$-hop risk propagation model, and discuss the new concept of propagation depth. Section \ref{sec:mul} presents the joint cyber risk of different types of nodes and the explicit formulas for computing the joint distribution of varying numbers of multi-type compromised nodes by using a novel backward elimination approach. Certain special cases are also presented in this section. In Section \ref{sec:param}, we study the effects of propagation depth and compromise probabilities on the joint cyber risk via stochastic ordering. We perform a simulation study in Section \ref{sec:simulation} to assess the joint cyber risk of a network system and obtain certain new insights. In the last Section, we conclude our main results and present some discussions. 	
	
	\section{ $L$-hop risk propagation model}\label{sec: propagation}
\subsection{Existing risk propagation models}
	There exists several risk propagation models in the literature, which aim to understand the propagation dynamics. For example, the well-known  SIS  and susceptible-infectious-recovered (SIR) epidemic models \cite{pastor2001epidemic} and their variations were widely studied in the areas of biology, epidemiology, and cyber security \cite{wang2014modeling,da2016peis,da2019modeling}. These models are used to describe the asymptotic steady status of an epidemic in a population or a virus over a computer network by employing certain stochastic process theories \cite{VanMieghemTON09}. For comprehensive reviews of SIS, SIR, and their variations, please refer to recent surveys \cite{pastor2001epidemic,wang2014modeling}. Another popular risk propagation model was originally proposed from the perspective of game theory \cite{kunreuther2003interdependent} and generalized to a probabilistic model \cite{Johnson2010}. The generalized model is named as the one-hop model and is further extended to the multi-hop model \cite{laszka2018assessment}.
Specifically, the one-hop model describes two types of risks encountered by a network. First is the direct attack, i.e., risk from outside the network, where a node is directly compromised by attacks, which includes the drive-by-download attacks (i.e., a node is compromised because its user visits a malicious website), or outside hacker attack (i.e., a node is directly compromised by the hacker). The second is the indirect attack, i.e., risk from within the network, where a node is compromised by attacks from its compromised neighbors. These two types of risks are also called the pull- and push-based risks, respectively, in the literature \cite{xu2015cyber}. In a one-hop model, the risk within the network only propagates one hop. Thus, a node only encounters risks from the outside or its immediate neighbors. Specifically, if a node $i$ is directly compromised by the risk from outside, it has the ability to propagate the risk to its healthy neighbor $j$ once with a probability $q_{ij}$. However, if the node is indirectly compromised, it cannot propagate the risk. The multi-hop model introduced in \cite{laszka2018assessment} allows a compromised node $i$ to propagate the risks irrespective of the node $i$ being directly or indirectly compromised. Therefore, a compromised node $i$ can compromise its healthy neighbor $j$ once with probability $q_{ij}$, irrespective of how the node $i$ is compromised.

\subsection{Proposed $L$-hop risk propagation model}	
In this section, we introduce the $L$-hop risk propagation model, i.e., the risk from a direct attack, is allowed to propagate $L$ hops within the network. For the purpose of illustration, a 2-hop risk propagation over a network is displayed in Figure \ref{fig:2-hop}. Nodes 1, 5, and 8 are directly compromised by the risks from outside the network. Those nodes propagate the risks to their neighbors, and compromise nodes 2, 4, and 9. The compromised nodes 2, 4, 9 further propagate the risks to their neighbors, and compromise nodes 7 and 12.
 \begin{figure}[hbtp!]
\centering
\includegraphics[width=.4\textwidth] {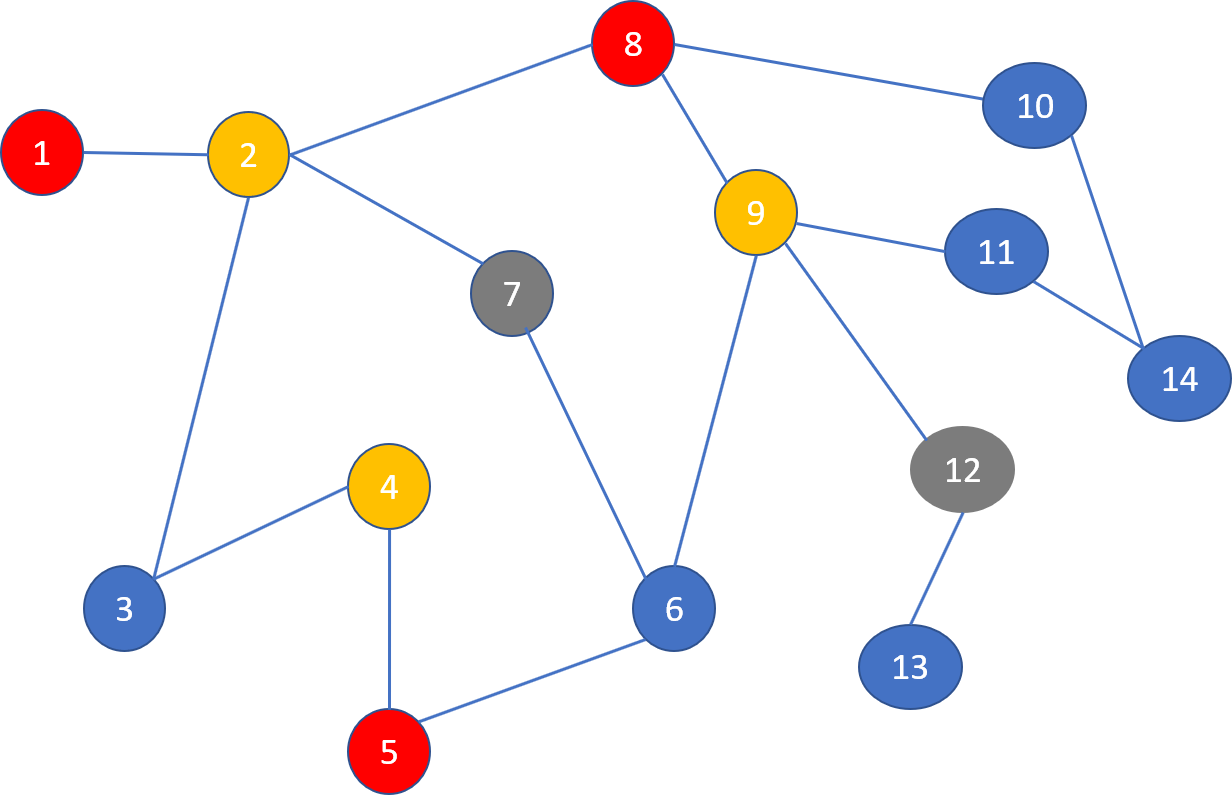}
\caption{A 2-hop risk propagation over a network. Blue color represents the healthy nodes. Red color represents the nodes (1, 5, and 8) that are compromised directly by the risk from outside the network. Gold color represents the nodes (2, 4, and 9) that are compromised by the directly-compromised nodes. Grey color represents the nodes (7 and 12) that are compromised by their infected neighbors. \label{fig:2-hop}}
\end{figure}

 {The following reasons  motivate us to introduce the $L$-hop risk propagation model.
\begin{itemize}
\item The propagation cannot be unlimited for most of cyber networks. Theoretically, the propagation depth is at most the length of the longest path between two nodes in the network. Hence, the $L$-hop model is a  natural extension of  the multi-hop model, with $L$ equal to the length of the longest path between two nodes in a network. Practically, the propagation will be stopped by the network defender after the compromise is detected on the network.
\item The propagation depth $L$ provides a new index for assessing the cyber risk of  network system.
  Specifically, the network defender can use this index to determine the propagation power of a new risk (e.g., a new malware). If $L$ is large, the new risk is considered to have more power for propagating over the network.  From the other hand,  this index can also be used for assessing the defense level of a company network. That is, if a less contagious malware can propagate more hops over a company network, the defense level of the network can be marked as low.  This would be particularly useful for the insurance company to perform the risk scoring.
  \item It should be noted that the $L$-hop risk propagation model is a nontrivial extension to the one-hop model as the propagation depth $L$ can significantly affect the dynamics of risk propagation. In \cite{laszka2018assessment}, it was argued that the $L$-hop model can be regarded as a certain one-hop model with a power matrix of indirect compromise probabilities between nodes. Unfortunately, this perspective is incorrect.  This is simply because that the  matrix of indirect compromise probabilities with certain power may not be  a probability matrix  (see   Section \ref{sec:mul} for a specific example).
  \end{itemize}
}
It is interesting to see that  the $L$-hop model can be connected to the $L$-hop percolation model in the literature of cyber security \cite{shang2011hop,shao2015percolation}. The $L$-hop percolation model can characterize the scenario of a network under attack by a computer malware and under the control of a bot master. The network defender can directly detect and delete certain bots (i.e., compromised nodes) and can trace the risk from the bots within their $L$-hop neighborhood and remove them \cite{shang2011hop}. Although both $L$-hop risk propagation and percolation models consider the risk propagation over $L$ hops, the proposed risk propagation focuses on the propagation dynamics from the perspective of risk management while the percolation model focuses on the strategies for removing the risk, i.e., from the perspective of network defense.

	\section{Joint cyber risk of a network system}\label{sec:mul}
	In this section, we analyze the joint cyber risk of heterogeneous nodes under the $L$-hop risk propagation model, namely, the multivariate distribution of numbers of multi-type  compromised nodes.
	
	We consider  an undirected finite network graph $G(V,E)$,
	where $V=\{1,2,\ldots,N\}$ is the  set of nodes with size $N=|V|$, and $E=\{e_{ij}: i,j \in V\}$ is the set of edges. 
	 We assume  that $G$ has $M$ different types of nodes, such as abstract computers, e-devices, and e-components. Let $\mathcal{S}_i$ be the set of all type $i$ nodes with $|\mathcal{S}_i|=N_i> 0$, $i=1,\ldots,M$, $\mathcal{S}=\bigcup_{i=1}^M \mathcal{S}_i $, and $\sum_{i=1}^MN_i=N$.
	
	 Let $\bm p$ be the direct compromise probability vector
	 $${\bm p}=(p_1,\ldots,p_N),$$
	  where $p_i$ represents the probability that node $i$ is compromised by the direct attack, $i=1,\ldots,N$. Let $Q$ be the indirect compromise probability matrix, i.e.,
	   $$Q=(q_{ij})\in [0,1]^{N\times N},$$
where element $q_{ij}$ denotes the probability that node $j$ is compromised by the indirect attack from node $i$, where $i,j=1,\ldots,N$, with $q_{ij}=0$ for $i=j$.
	
{In the subsequent discussion, we assume that all random compromise events are statistically independent. We denote $\bar a=1-a$ for any $a\in \mathsf{R}$; $\bar A$ is the complement of any set $A$; further, $A\backslash B$ is the difference set of $A$ and $B$. The following conventional notations are also used throughout the paper.
	$$
	0^0=1,\ \sum_{\emptyset}=0,\ \prod_{\emptyset}=1,\binom{n}{m}=0 \ \mbox{for $m<0$ or $m>n$}.
	$$

\subsection{Backward elimination approach for a general case}
	Let $X_i$ be the number of compromised nodes in $\mathcal{S}_i$ for $i=1,\ldots,M$. Subsequently, we discuss the procedure to compute the joint probability of varying numbers of compromised nodes. To facilitate the discussion, we denote
$$
E_U=\{e_{ij}\in E:i,j\in U\},\quad U\subset V.
$$
 {Let $R_{G(U)}(C,D;L)$ }denote the conditional probability that only all nodes of $C$ are compromised provided that all nodes of $D$ are compromised directly over the network $G(U,E_U)$, where $D\subset C\subset U$.


	\begin{The} \label{th_gen}\rm For $L\ge 1$, the joint probability mass function of random vector $(X_{1},\ldots,X_{M})$ is,
	\begin{eqnarray}\label{eq:main}
		&& f(x_1,\ldots,x_M) \nonumber \\&=&\sum_{\substack{C_{1}\subset \mathcal{S}_{1}\\|C_{1}|=x_{1}}}\cdots\sum_{\substack{C_{M}\subset \mathcal{S}_{M}\\|C_{M}|=x_{M}}}\sum_{{D}_0\subset \mathcal{C}} \nonumber  \\
  &&\left\{\prod_{j\in {D}_0} p_j \cdot \prod_{i\in \overline{{D}_0}} \bar p_i\cdot R_{G(V)}(\mathcal{C},{D}_0;L)\right\},
		\end{eqnarray}
	where $0\le x_{i}\le N_{i}$, $\mathcal{C}=\cup_{i=1}^MC_i$, and
\begin{itemize}
\item[a)] for $L=1$,
\begin{eqnarray}\label{L1}
&&R_{G(V)}(\mathcal{C}, D_0;1)\nonumber\\
&=&\prod_{i\in \mathcal{C}\backslash D_0}\left(1-\prod_{j\in D_0}\bar q_{ji}\right)\prod_{v\in D_0, l\in \overline{\mathcal{C}}}\bar q_{vl};
\end{eqnarray}
\item[b)] for $L\ge 2$,
\begin{eqnarray}\label{eq:rec}
 && R_{G(V)}(\mathcal{C},{D}_0;L)\nonumber\\
 &&=\sum_{\substack{D_{j}\subset{\mathcal{C}\backslash\mathfrak{D}_{j-1}}\\j=1,\ldots,L-1}}
 \prod_{i=0}^{L-2}R_{G(V\backslash\mathfrak{D}_{i-1})}(D_i\cup D_{i+1},D_i;1)\nonumber\\
 &&\cdot R_{G(V\backslash \mathfrak{D}_{L-2})}(\mathcal{C}\backslash \mathfrak{D}_{L-2},D_{L-1};1)
\end{eqnarray}
with $\mathfrak{D}_k=\cup_{l=0}^kD_l$ for $k=0,\ldots,L-1$, and $\mathfrak{D}_{-1}=\emptyset$.
\end{itemize}
	\end{The}

	 \proof We define  the events as:
	 \begin{equation*}\label{eq:a1star}
	B_{C}=\left\{\mbox{Only nodes in }\ C\ \mbox{are compromised for $L$-hop propagation}\right\}
	\end{equation*}
	and
	\begin{equation*}\label{eq:a2star}
	A_{D}=\left\{\mbox{Only nodes in}\ D\ \mbox{are directly compromised}\right\}
	\end{equation*}
	for $C,D\subset V$.

	 When $0\le x_{i}\le N_{i}, i=1,\ldots,M$,
	\begin{eqnarray*}
	\p\left(X_{1}=x_{1},\ldots,X_{M}=x_{M}\right)=\sum_{\substack{C_{1}\subset \mathcal{S}_{1}\\|C_{1}|=x_{1}}}\cdots\sum_{\substack{C_{M}\subset \mathcal{S}_{M}\\|C_{M}|=x_{M}}}\p\left(B_{\mathcal{C}}\right),
	\end{eqnarray*}
	where $C_{i}$ represents the set of type $i$ nodes with the cardinality $x_{i}$, and $\mathcal{C}=\cup_{i=1}^M C_i$. Now, we consider  whether the nodes are directly compromised. Considering the condition on the directly compromised nodes,
	\begin{eqnarray*}
	\p\left(B_{\mathcal{C}}\right)=\sum_{D_0\subset\mathcal{C}}\p\left(B_{\mathcal{C}}|A_{D_0}\right)\p\left(A_{D_0}\right),
	\end{eqnarray*}
where
	\begin{eqnarray*}
		\p\left(A_{D_0}\right)=\prod_{j\in D_0} p_j\cdot\prod_{i\in \overline{D_0}}\bar p_i.
	\end{eqnarray*}

 It is to be noted that
$$
\p(B_{\mathcal{C}}|A_{{D_0}})=R_{G(V)}\left(\mathcal{C},{D_0};L\right).
$$
For $L=1$, only the propagation from the directly compromised nodes to their neighbors is permitted. Therefore, the probability that the nodes in $ \mathcal{C}\backslash{D}_0 $ are compromised indirectly by ${D}_0$ is
	$$
	\prod_{i\in \mathcal{C}\backslash {D_0}}\left(1-\prod_{j\in {D_0}}\bar q_{ji}\right),
	$$
	and the probability that the nodes in $\overline{\mathcal{C}}$ are not compromised by ${D_0}$ is
	$
	\prod_{v\in{D_0}, l\in \overline{\mathcal{C}}}\bar q_{vl}.
	$
Thus,
\begin{equation}\label{eq:p_theta}
R_{G(V)}(\mathcal{C},D_0;1)=
 \prod_{l\in \mathcal{C}\backslash D_0}(1-\prod_{j\in D_0}\bar q_{jl})\prod_{i\in D_0,h\in\overline{\mathcal{C}}}\bar q_{ih}.
 \end{equation}

For $L\ge 2$, the computation of the key component $\p(B_{\mathcal{C}}|A_{\mathcal{D}})$ is significantly complex. To overcome the computational complexity, we introduce  the following novel {\em backward elimination} approach for computing the joint probability mass function.

The idea of backward elimination is motivated from the perspective of dynamic propagation. Thus,
we consider the $L$-hop propagation as $L$ rounds of propagation.
We define  $B_{C}^{(m)}$ as the event that
only the nodes of $C$ are compromised after $m$ rounds of propagations for $1\le m\le L$ and $C\subset V$. Note that $B_{C}^{(L)}\equiv B_C$.

Let us consider the first round of propagation.
By conditioning on the exact number of compromised nodes in $\mathcal{C}\backslash{D_0}$ after the first round of propagation, by the law of total probability, the following is satisfied:
\begin{eqnarray}\label{rec}
&& R_{G(V)}(\mathcal{C},{D_0};L)\nonumber\\ &=&\p\left(B_{\mathcal{C}}^{(L)}|A_{D_0}\right)\nonumber\\
&=&\sum_{D_1\subset\mathcal{C}\backslash {D_0}}\p\left(B_{\mathcal{C}}^{(L)}|B^{(1)}_{D_1},A_{D_0}\right)\p\left(B_{D_1}^{(1)}|A_{D_0}\right),
\end{eqnarray}
where
\begin{equation}\label{eq:round}
\p\left(B_{D_1}^{(1)}|A_{D_0}\right)
=R_{G(V)}(D_1\cup{D_0},D_0;1).
\end{equation}

The conditional probability
$$
\p\left(B_{\mathcal{C}}^{(L)}|B^{(1)}_{D_1},A_{D_0}\right)
$$
can be efficiently computed by using the backward elimination approach as follows. After the first round of propagation, nodes in ${D_0}$ and all the edges from ${D_0}$ to $V\backslash {D_0}$ are eliminated as those elements do not play a role in the second round of propagation. Therefore, the network $G(V)$ reduces to a subnetwork $G(V\backslash {D_0})$. Given $B_{D_1}^{(1)}$ (i.e., only the nodes in $D_1\in\mathcal{C}\backslash {D_0}$ are compromised in the first round), the event that the nodes in $\mathcal{C}$ are compromised owing to $A_{D_0}$ after $L$ rounds of propagations is equivalent to that of
the nodes in $\mathcal{C}\backslash{D_0}$ being compromised by the nodes in $D_1$ over the network $G(V\backslash{D_0})$ after $L-1$ rounds of propagations (namely, $B^{(L-1)}_{\mathcal{C}\backslash{D_0}}|A_{D_1}$). Thus,
\begin{equation}\label{L-1}
\p\left(B_{\mathcal{C}}^{(L)}|B^{(1)}_{D_1},A_{D_0}\right)=R_{G(V\backslash {D_0})}(\mathcal{C}\backslash {D_0},D_1;L-1).
\end{equation}
Substituting \eqref{eq:round} and \eqref{L-1} into \eqref{rec} yields
\begin{eqnarray}\label{iter}
&& R_{G(V)}(\mathcal{C},D_0;L)\nonumber \\
&=&\sum_{D_1\subset{\mathcal{C}\backslash D_0}} R_{G(V)}(D_1\cup{D_0},D_0;1)
\nonumber\\&&\cdot R_{G(V\backslash {D_0})}(\mathcal{C}\backslash D_0,D_1;L-1).
\end{eqnarray}
After $L-1$ iterations of \eqref{iter}, the explicit expression can be written as:

\begin{eqnarray*}\label{eq:recursive}
&& R_{G(V)}(\mathcal{C},{D}_0;L)\\
 &=&\sum_{\substack{D_{1}\subset{\mathcal{C}\backslash\mathfrak{D}_{0}}}}\sum_{\substack{D_{2}\subset{\mathcal{C}\backslash\mathfrak{D}_{1}}}}\cdots \sum_{\substack{D_{L-1}\subset{\mathcal{C}\backslash\mathfrak{D}_{L-2}}}}\\
 &&\prod_{i=0}^{L-2} R_{G(V\backslash\mathfrak{D}_{i-1})}(D_i\cup D_{i+1},D_i;1)\\
 &&\cdot R_{G(V\backslash \mathfrak{D}_{L-2})}(\mathcal{C}\backslash \mathfrak{D}_{L-2},D_{L-1};1),
\end{eqnarray*}
where $\mathfrak{D}_k=\cup_{l=0}^kD_l$ for $k=0,\ldots,L-1$, and $\mathfrak{D}_{-1}=\emptyset$.

The required result is presented subsequently. \qed

Theorem \ref{th_gen} provides explicit formulas for computing the joint probability for the varying numbers of multi-type compromised nodes with a propagation depth $L$. It should be emphasized that the primary expression in Eq. \eqref{eq:rec} only considers the case of $L=1$, which is readily satisfied in Eq. \eqref{L1}. Hence, the joint probability can be effectively computed from Eq. \eqref{eq:main}. In particular, Theorem \ref{th_gen} provides an explicit formula for a special case of a multi-hop model (i.e., $L$ is the length of the largest path of a network).

The key idea for deriving the explicit expressions for the joint probability is the proposed backward elimination approach. To further explain the idea of the backward elimination approach, we consider  the two-hop risk propagation illustrated in Figure \ref{fig:2-hop}. At the initial stage (i.e., stage 0), we assume that the network only includes the directly compromised nodes (i.e., 1, 5, 8), as illustrated in Figure \ref{fig:d0}. Let us assume that these compromised nodes indirectly compromise nodes 2, 4, 9 successfully at stage 1. The idea of backward elimination is to remove nodes 1, 5, 8, and the edges $\{e_{12},e_{54},e_{56},e_{82},e_{89},e_{810}\}$ as those nodes and edges do not play a role in the subsequent propagation. Then, we obtain a smaller network, as shown in Figure \ref{fig:d1}. As the propagation depth is $L=2$, nodes 2, 4, and 9 further compromise nodes 7 and 12 at stage 2. Now, we can remove nodes 2, 4, 9, and the edges $\{e_{27},e_{23},e_{43},e_{96},e_{912},e_{911}\}$, which results in a much smaller network, as illustrated in Figure \ref{fig:d2}. 	
\begin{figure*}[htb!]
\centering
\subfigure[Stage 0]{\includegraphics[width=0.4\textwidth]{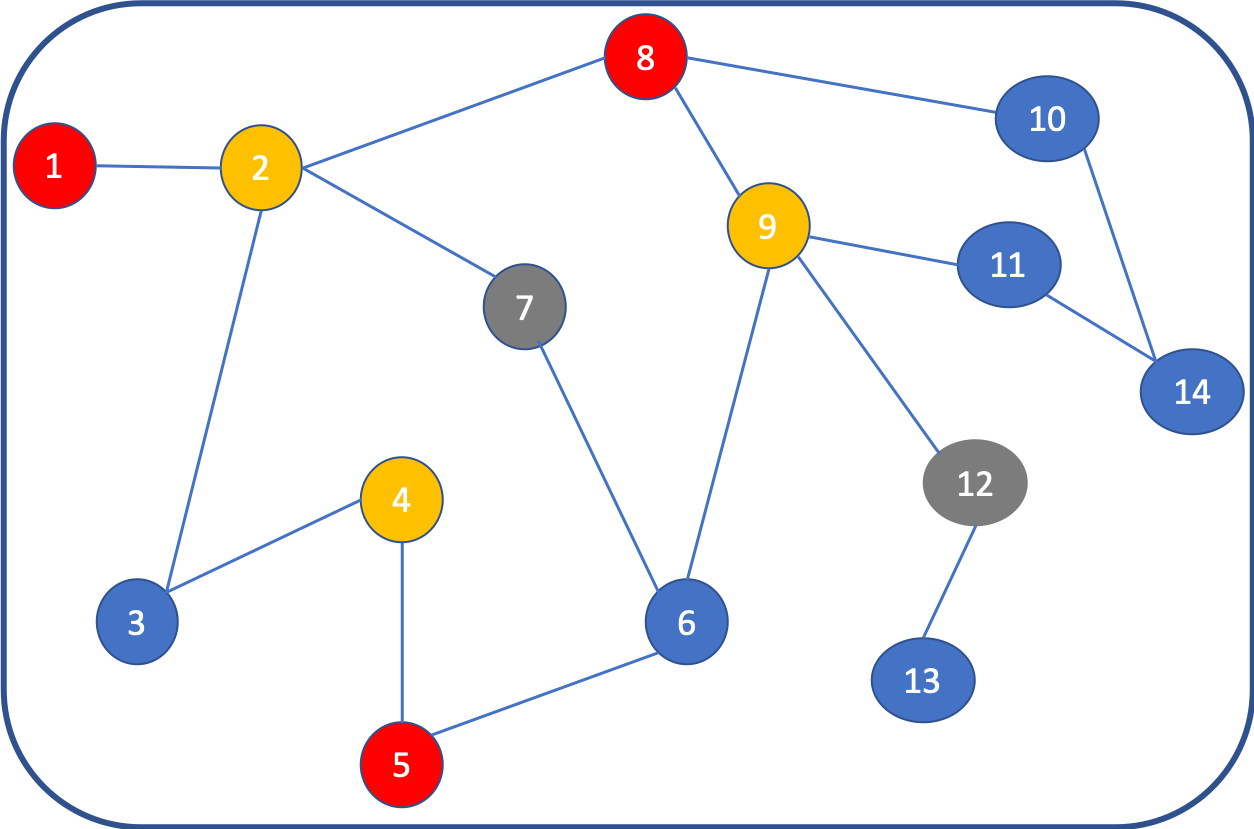}\label{fig:d0}}\\
 \subfigure[Stage 1]{\includegraphics[width=0.4\textwidth]{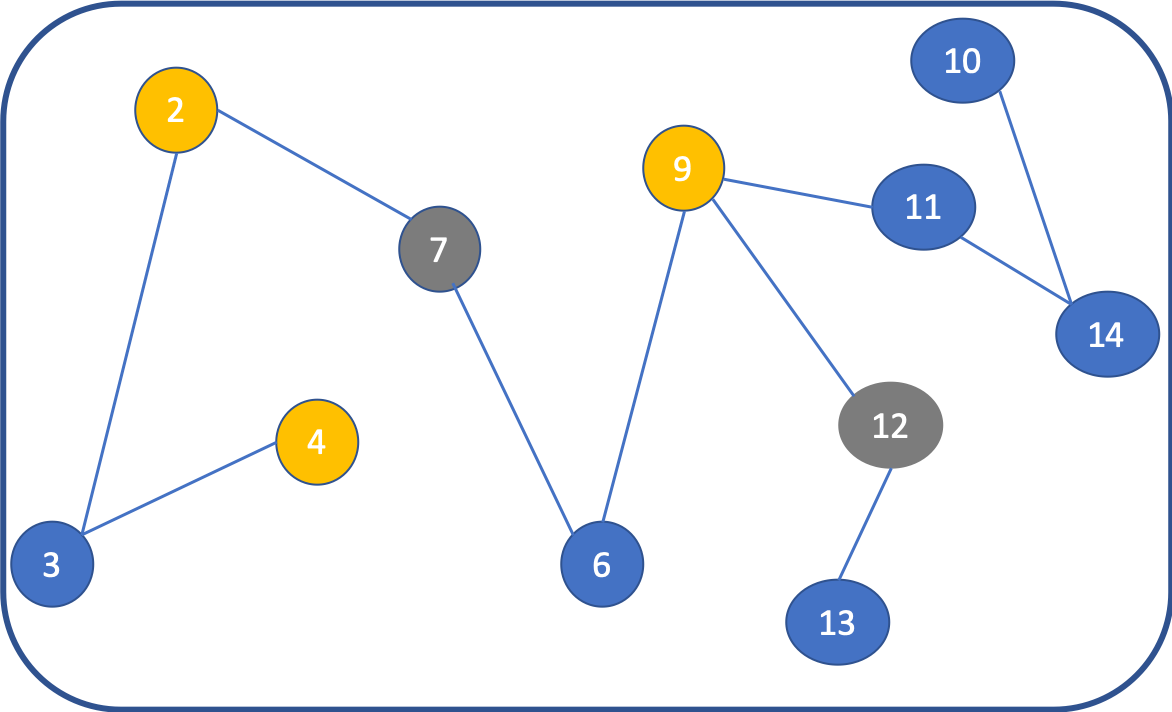}\label{fig:d1}}\qquad
 \subfigure[Stage 2]{\includegraphics[width=0.4\textwidth]{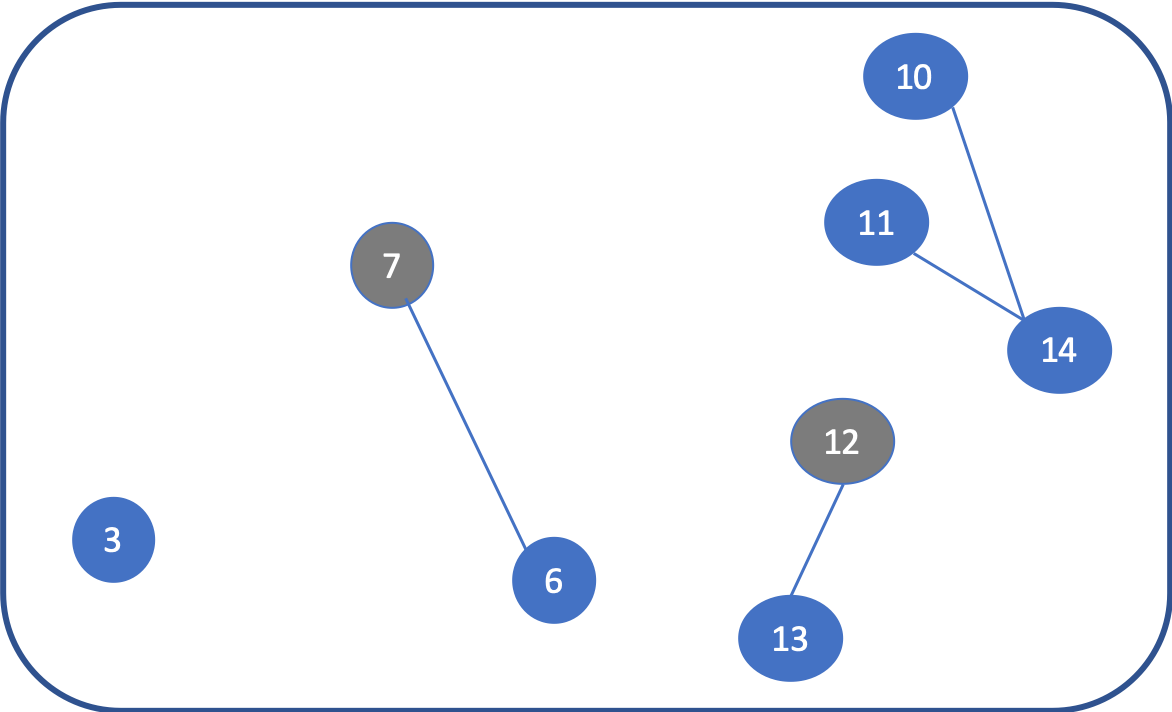}\label{fig:d2}}
 \caption{Illustration of the backward elimination approach\label{fig:elimination}}
\end{figure*}

 {  In \cite{laszka2018assessment}, it was stated that the $L$-hop model with indirect compromise probability matrix  $Q$  can be regarded as the one-hop model with  indirect compromise probability matrix $Q^L$.  Unfortunately, this statement is incorrect as $Q^L$ may not even be  a probability matrix. For example, consider the homogeneous model with $N=4$, $L=2$, $q_{ij}=0.8$ for  $1\le i\neq j\le 4$, and $q_{ii}=0$ for  $i=1,2,3,4.$ It can be seen that all of off-diagonal components of $Q^2$ are 1.28, which means that $Q^2$ can not be a matrix of indirect compromise probabilities.
}

{
 \subsection{Special cases}	
 In this section, we discuss certain specific network topologies for which the multivariate distributions of varying numbers of compromised nodes have simpler forms.
\subsubsection{Complete network} Consider a complete graph network consisting of $M$ types of nodes with $\sum_{i=1}^MN_i=N$. We further assume  that the compromise probabilities are homogeneous; therefore,
\begin{equation}\label{homo}
p_l=p, \quad q_{ij}=q,
\end{equation}
for $l\in\{1,\ldots,N\}$ and $i\neq j\in\{1,\ldots,N\}$.

Note that the conditional probability $R_{G(U)}(C,D;L)$ now only depends on $|C|$ and $|D|$ for this case. To simplify the notations, we define
$R_{K_u}(c,d;L)$ as the conditional probability that only certain $c-d$ nodes are compromised indirectly, provided that certain $d$ nodes are compromised directly over a complete network $K_u$ with $u$ nodes, where $d\le c\le u$.
Owing to the symmetry, Eqs. \eqref{L1}, \eqref{iter}, and \eqref{eq:rec} can be simplified as
\begin{equation}\label{homo_form1}
R_{K_u}(c,d;1)=(1-\bar q^d)^{c-d} \bar q^{d(u-c)},
\end{equation}
\begin{eqnarray}\label{homo_form}
R_{K_u}(c,d;L)&=&\sum_{i=0}^{c-d}\binom{c-d}{i}R_{K_u}(d+i,d;1)\nonumber\\
&&\cdot R_{K_{u-d}}(c-d,i;L-1),
\end{eqnarray}
and
\begin{eqnarray}\label{homo_ex}
&& R_{K_u}(c,d;L)\nonumber\\&=&\sum_{d_1=0}^{c-d}\sum_{d_2=0}^{c-\delta_1}\cdots\sum_{d_{L-1}=0}^{c-\delta_{L-2}}\prod_{j=0}^{L-2}
\binom{c-\delta_j}{d_{j+1}}(1-\bar q^{d_j})^{d_{j+1}}\nonumber\\
&&\cdot\bar q^{d_j(u-\delta_{j+1})} (1-\bar q^{d_{L-1}})^{c-\delta_{L-1}}\bar q^{d_{L-1}(u-c)}
\end{eqnarray}
with $\delta_k=d+\sum_{i=1}^kd_i$ for $k=0,\ldots,L-1$ and $\delta_{-1}=0$.

Then, we derive  the following result for a network with complete topology and homogeneous compromise probabilities.
\begin{Pro}\rm\label{homo_cor}
Under the assumption \eqref{homo}, for a given $L\ge 1$, the following equation is satisfied:
\begin{eqnarray*}
f(x_1,\ldots,x_M)=\prod_{i=1}^M\binom{N_i}{x_i} \sum_{d=0}^{\chi}\binom{\chi}{d}\left[p^{d}\bar p^{N-d}R_{K_N}\left(\chi,d;L\right)\right]
\end{eqnarray*}
with $\chi=\sum_{i=1}^Mx_i$, where $R_{K_N}\left(\chi,d;L\right)$ can be computed by \eqref{homo_form1} and \eqref{homo_ex} for $L=1$ and $L\ge 2$, respectively.
\end{Pro}

\subsubsection{Star network} Consider a star network with one hub and $N-1$ leaves. The nodes can be divided into two categories: hub (type I) and leaves (type II), i.e., $M=2$. We define the compromise probability as follows. Let $p_I (p_{II})$ denote the probability of direct compromise of the hub (a leaf), and $q_{I,II}$ ($q_{II,I}$) denote the probability that a leaf (the hub) is compromised by the hub (a leaf). Note that for a star network, the propagation depth $L$ is at most two. From Theorem \ref{th_gen}, we have the following closed forms of the joint mass functions for the joint risk.
\begin{Pro}\rm For a star network with the aforementioned two types of nodes, we have:
\begin{itemize}
\item[a)] for $L=1$, the joint mass function of $(X_1,X_2)$ is given by
\begin{equation*}
	f(0,m)=\binom{N-1}{m}p_{II}^m\bar p_{I}\bar p_{II}^{N-1-m}\bar q_{II,I}^m
	\end{equation*}
	and
	\begin{eqnarray*}
&&f(1,m)\\
&=&\binom{N-1}{m}p_{II}^m\bar p_{I}\bar p_{II}^{N-1-m}(1-\bar q_{II,I}^m)\\
&&+\binom{N-1}{m}\sum_{d=0}^m\binom{m}{d}p_{I}p_{II}^d\bar p_{II}^{N-1-d}q_{I,II}^{m-d}\bar q_{I,II}^{N-1-m};
	\end{eqnarray*}
\item[b)] for $L=2$, the joint mass function of $(X_1,X_2)$ is given by
$$
f(0,m)=\binom{N-1}{m} \bar p_I p_{II}^m\bar p_{II}^{N-1-m}\bar q_{II,I}^m;
$$
moreover, for $0\le m\le N-1$,
\begin{eqnarray*}
&&f(1,m)=\binom{N-1}{m}\\
&&\cdot \left(\sum_{d=1}^m\binom{m}{d}\bar p_I p_{II}^d\bar p_{II}^{N-1-d}(1-\bar q_{II,I}^d)\bar q_{I,II}^{N-1-m}q_{I,II}^{m-d}\right.\\
&&  \left.+\sum_{d=0}^m\binom{m}{d}p_I p_{II}^d\bar p_{II}^{N-1-d} \bar q_{I,II}^{N-1-m}q_{I,II}^{m-d}\right).
\end{eqnarray*}
\end{itemize}
\end{Pro}

\subsubsection{Complete bi-partite network} A complete bi-partite graph $K_{N_1,N_2}$ consists of two disjoint sets $\mathcal{S}_1$ and $\mathcal{S}_2$ containing corresponding $N_1$ and $N_2$ nodes, such that all nodes in $\mathcal{S}_1$ are connected to all
	nodes in $\mathcal{S}_2$, while within each set there are no connections. 
	
	We assume  that the direct compromise probabilities of nodes in each set are the same, i.e., $p_i\equiv p_{I}$ and $p_j\equiv p_{II}$, for all $i\in \mathcal{S}_1$ and $j\in \mathcal{S}_2$, and the propagation probability from any node in $\mathcal{S}_1$ ($\mathcal{S}_2$) to any node in $\mathcal{S}_2$ ($\mathcal{S}_1$) is the same, denoted by $q_{I,II}$ ($q_{II,I}$). Then, for the one-hop prorogation, we derive  the subsequent result.
	
	\begin{Pro}\label{corbi} \rm Assume a complete bi-partite network with disjoint sets $\mathcal{S}_1$ and $\mathcal{S}_2$. For propagation depth $L=1$, the joint probability mass function of $(X_1,X_2)$ is given by
	\begin{eqnarray*}\label{eqbi}
	&&f(m_1,m_2)\\
&=&\binom{N_1}{m_1}\binom{N_2}{m_2}\sum_{d_1=0}^{m_1}\sum_{d_2=0}^{m_2}\left\{\binom{m_1}{d_1}\binom{m_2}{d_2}p_{I}^{d_1} p_{II}^{d_2}\right.\\
&&\cdot(1-p_{I})^{N_1-d_1}(1-p_{II})^{N_2-d_2} \\
	&&\left.\cdot\left[1-(1-q_{I,II})^{d_1}\right]^{m_2-d_2}\cdot \left[1-(1-q_{II,I})^{d_2}\right]^{m_1-d_1}\right.\\\notag
	&&\left.\cdot(1-q_{I,II})^{d_1(N_2-m_2)}\cdot \left(1-q_{II,I}\right)^{d_2(N_1-m_1)}\right\},
	\end{eqnarray*}
	for $0\le m_1\le N_1$ and $0\le m_2\le N_2$.
	 \end{Pro}


To conclude this section, we present an example illustrating the backward elimination approach. For simplicity, we consider  the case of a complete network with homogeneous compromise probabilities.
	
	 \begin{Exa}\label{example1}\rm
 Consider a complete graph network with $N=5$ nodes. Assume that there are two types of nodes: $S_1=\{1,2\}$ and $S_2=\{3,4,5\}$; further, the propagation depth is $L=2$. We assume  that for $1\le i,j\le 5$,
$$
p_i=p, \quad q_{ij}=q.
$$
Subsequently, we compute the joint mass function of $(X_1,X_2)$. First, it is easy to see that
$$
f(0,0)=\bar p^5.
$$
Further, by Proposition \ref{homo_cor}, we derive
$$
f(1,0)=2\times\left[\bar p^5R_{K_5}(1,0;L)+p\bar p^4R_{K_5}(1,1;L)\right].
$$
Note that the indirect compromise probability is homogeneous; thus, we obtain
$$
R_{K_5}(1,0;2)=0, \quad
R_{K_5}(1,1;2)=\bar q^4.
$$
Then, we calculate that
$$
f(1,0)=2p\bar p^4\bar q^4.
$$
Similarly, we have
$$
f(0,1)=3p\bar p^4\bar q^4.
$$
To maintain conciseness, we only compute  the expression of $f(1,1)$ for the illustration. Using Proposition \ref{homo_cor}, we derive  the following expression
\begin{eqnarray}\label{ss3}
f(1,1)=6\times\left[2p\bar p^4R_{K_5}(2,1;2)+p^2\bar p^3R_{K_5}(2,2;2)\right].
\end{eqnarray}
 Applying Eq. \eqref{homo_ex}, it follows that
$$
R_{K_5}(2,1;2)=q\bar q^4R_{K_4}(1,0;1)+q\bar q^3R_{K_4}(1,1;1)
$$
and
$$
R_{K_5}(2,2;2)=\bar q^6R_{K_3}(0,0;1).
$$
Using Eq. \eqref{homo_form1}, we also calculate:
$$
R_{K_4}(1,0;1)=0, \quad
R_{K_4}(1,1;1)=\bar q^3,
$$
and
$$
R_{K_3}(0,0;1)=1.
$$
Therefore,
\begin{eqnarray}
R_{K_5}(2,1;2)&=&q\bar q^6,\label{ss1}\\
R_{K_5}(2,2;2)&=&\bar q^6.\label{ss2}
\end{eqnarray}
Substituting \eqref{ss1} and \eqref{ss2} into \eqref{ss3} yields
$$
f(1,1)=6\times\left(2p\bar p^4q\bar q^6+p^2\bar p^3\bar q^6\right).
$$
The other cases can be computed similarly.

Table~\ref{tab4} lists the joint probabilities of $(X_1,X_2)$ for $p=0.2$ and $q=0.1$ with $L=2,3,4$ (note that $L=4$ demonstrates a multi-hop model).
\begin{table}[hbtp!]
	\centering
	\caption{  Joint probability for $p=0.2$, $q=0.1$, and $L=2,3, 4$.	\label{tab4}}

	\begin{tabular}{c|cccc|c}
		\hline\hline
		$X_1 \diagdown X_2$ & 0 & 1 & 2 & 3 &Total\\
		\hline
	    \multicolumn{6}{c}{$L=2$}\\	\hline
		0 & 0.3277  &  0.1612 & 0.0588 & 0.0131 &0.5608\\
		1 & 0.1075  &  0.1175 & 0.0788 & 0.0245&0.3283\\	
		2 & 0.0196  &  0.0394 & 0.0367 & 0.0152&0.1109\\
		\hline
		Total&0.4548&0.3181&0.1743&0.0528& \\ \hline
 	    \multicolumn{6}{c}{$L=3$}\\	\hline
			0 & 0.3277  &  0.1612 & 0.0588 & 0.0126&0.5603\\
		1 & 0.1075  &  0.1175 & 0.0755 & 0.0255& 0.326\\
		2 & 0.0196  &  0.0377 & 0.0383 & 0.0181&0.1137\\
		\hline
		Total&0.4548&0.3164&0.1726 &0.0562 & \\ \hline
	    \multicolumn{6}{c}{$L=4$}\\	\hline
		0 & 0.3277  &  0.1612 & 0.0588 & 0.0126&0.5603\\
		1 & 0.1075  &  0.1175 & 0.0755 & 0.0253&0.3258\\
		2 & 0.0196  &  0.0377 & 0.0380 & 0.0186&0.1139\\
		\hline
		Total&0.4548&0.3164&0.1723&0.0565 \\
		\hline\hline	
	\end{tabular}
\end{table}
It can be observed that when $L$ increases, the joint probability of compromising more nodes demonstrates an overall increasing trend. For example, when $L=2$, we observe $P(X_1=2, X_2=3)=0.0152$, which increases to $0.0181$ for $L=3$, and to $ 0.0186$ for $L=4$. In particular, the probability of $P(X_1=2)$ increases from $0.1109$ for $L=2$ to $0.1139$ for $L=4$,
and the probability of $P(X_2=3)$ increases from $0.0528$ for $L=2$ to $0.0565$ for $L=4$. This observation is confirmed by the theoretical results obtained in Section \ref{sec:param}.
\end{Exa}
	
 }	\section{Effects of propagation parameters}\label{sec:param}
	In this section, we discuss how the propagation depth $L$, direct infection probability vector $\bm p$, and indirect {compromise} probability matrix $Q$ affect the joint distribution of the number of compromised nodes. Intuitively, the joint risk of nodes over the network should increase with increasing values of $L$, $\bm p$, and $Q$.  In the following, we provide rigorous mathematical proofs to verify those intuitions via the tool of stochastic orders, which has been widely used in statistics, operation research, risk management, and several other areas \cite{navarro2015orderings,shaked2007stochastic}.

We first recall the following multivariate stochastic order \cite{shaked2007stochastic}.

	\begin{Def}\rm~ Suppose $\textbf{X}=(X_{1},\ldots,X_{n})$ and $\textbf{Y}=(Y_{1},\ldots,Y_{n})$ are two random vectors. Then, random vector $\textbf{X}$ is said to be smaller than $\textbf{Y}$ in the usual multivariate stochastic order (denoted by $\textbf{X}\leq_{\rm st}\textbf{Y}$) if $\mathbb{E}[\phi(\textbf{X})]\leq\mathbb{E}[\phi(\textbf{Y})]$ for all nondecreasing functions $\phi$.
	\end{Def}
It should be noted that \cite{shaked2007stochastic}:
	$$\textbf{X}\leq_{\rm st}\textbf{Y}\Rightarrow \p(X_1\le x_1,\ldots,X_n\le x_n)\ge \p(Y_1\le x_1,\ldots,Y_n\le x_n)$$
	and
	$$\textbf{X}\leq_{\rm st}\textbf{Y}\Rightarrow \p(X_1> x_1,\ldots,X_n>x_n)\le \p(Y_1> x_1,\ldots,Y_n> x_n).$$
	
	Let $(X_1,\ldots,X_M|L,\bm p,Q)$ be the number vector of the compromised nodes for the given parameters $L$, $\bm p$, and $Q$. The following result shows that if the propagation depth $L$ is larger, there are more compromised nodes in the sense of the multivariate stochastic order.

	\begin{The}\rm We assume  a network has $M$ types of nodes, with a propagation depth $L$, direct compromise probability vector $\bm p$, and indirect {compromise} probability matrix $Q$. Then, the random vector $(X_1,\ldots,X_M)$ increases with increasing values of $L$ in the sense of multivariate stochastic order, i.e., for $L\le L'$,
	\begin{equation*}\label{ineq1}
	\left(X_1,\ldots,X_M|L,\bm p,Q\right)\le_{\rm st}\left(X_1,\ldots,X_M|L',\bm p,Q\right);
	\end{equation*}
and hence,
$$\left[X_i|L,\bm p,Q\right]\le_{\rm st}\left[X_i|L',\bm p,Q\right],\, i=1,\ldots,M,$$ and 	
$$\left[\sum_{i=1}^M X_i\bigg|L,\bm p,Q\right]\le_{\rm st}\left[\sum_{i=1}^M X_i\bigg|L',\bm p,Q\right].$$	
	 \end{The}
	 \proof Note that for the network under cyber threats, the set of compromised nodes in the $L$-hop model is always a subset
	 of that in the $(L+1)$-hop model. This is because the propagation is allowed to propagate one more step. Therefore,
	 $$
	\left(X_1,\ldots,X_M|L,\bm p,Q\right)\le_{\rm a.s.}\left(X_1,\ldots,X_M|L+1,\bm p,Q\right),
	$$
	where a.s. represents ``almost surely". According to Theorem 6.B.1 in \cite{shaked2007stochastic},
	 $$
	\left(X_1,\ldots,X_M|L,\bm p,Q\right)\le_{\rm st}\left(X_1,\ldots,X_M|L+1,\bm p,Q\right),
	$$
	which further implies
	$$\left[X_i|L,\bm p,Q\right]\le_{\rm st}\left[X_i|L',\bm p,Q\right],\, i=1,\ldots,M,$$
	and
	$$\left[\sum_{i=1}^M X_i\bigg|L,\bm p,Q\right]\le_{\rm st}\left[\sum_{i=1}^M X_i\bigg|L',\bm p,Q\right].$$ \qed
	{

 Theorem \ref{ineq1} verifies the intuition that if the risk has more power to propagate, i.e., the propagation depth $L$ is larger, then more nodes over the network will be compromised.
 It further shows that the number of compromised nodes is always larger when $L$ is larger for each type. In particular, for $L\le L'$,
\begin{eqnarray*}
&&\p(X_1\le x_1,\ldots,X_M\le x_m|L,\bm p,Q)\\
&\ge& \p(X_1\le x_1,\ldots,X_M\le x_m|L',\bm p,Q).
\end{eqnarray*}
 This implies that the joint distribution function of the number of compromised nodes decreases when the propagation depth increases. Similarly, for $L\le L'$,
\begin{eqnarray*}
&&\p(X_1> x_1,\ldots,X_M> x_m|L,\bm p,Q)\\
&\ge& \p(X_1> x_1,\ldots,X_M> x_m|L',\bm p,Q).\end{eqnarray*}
This indicates that the joint survival function of the number of compromised nodes increases (i.e., a more sever network environment) when the propagation depth increases. {These results coincide with the findings listed in Table \ref{tab4}.}

}
	Subsequently, we analyze how the direct and indirect probabilities affect the number of compromised nodes.
	\begin{The}\rm \label{thm:infection} We assume  a network has $M$ types of nodes, with a propagation depth $L$, direct compromise probability vector $\bm p$, and indirect {compromise} probability matrix $Q$. Then, the random vector $(X_1,\ldots,X_M)$ increases when:
	\begin{itemize}
	  \item [a)] $\bm p$ increases in the sense of the multivariate stochastic order, i.e., for $\bm p\le \bm p'$,
	      \begin{equation*}\label{ineq2}
	(X_1,\ldots,X_M|L,\bm p,Q)\le_{\rm st}(X_1,\ldots,X_M|L,\bm p',Q);
	\end{equation*}
	  \item [b)] $Q$ increases in the sense of the multivariate stochastic order, i.e., for $Q\le Q'$,
	      \begin{equation*}\label{ineq3}
	(X_1,\ldots,X_M|L,\bm p,Q)\le_{\rm st}(X_1,\ldots,X_M|L,\bm p,Q').
	\end{equation*}
	\end{itemize}
	\end{The}
	
	 \proof The proof is determined by using the coupling method in probability theory \cite{lindvall2002lectures}.
	
	a) Considering a network $A$, let $I_j$ be a Bernoulli random variable with $I_{j,A}=1$, which represents that node $j$ is compromised successfully by a direct attack; then,
	$$\p(I_{j,A}=1)=p_j, \, j=1,\ldots,N.$$
	Now, we construct   a new network $B$ with the same types of nodes, a propagation depth $L$, and indirect {compromise} probability matrix $Q$. We assume  that the new network B has a different direct compromise probability vector $\bm p'\ge \bm p$. The vector $\bm p'$ is constructed as follows. Let $I_{j,B}$ represent a node $j$ that is successfully  compromised by a direct attack over network $B$. We define  that if $I_{j,A}=1$, then $I_{j,B}=1$; if $I_{j,A}=0$, then $I_{j,B}=1$ with probability ${(p_j'-p_j)/}(1-p_j)$. Then,
	$$\p(I_{j,B}=1)=p'_j.$$
	Therefore,
	$$
	\left(X_1^{A},\ldots,X_M^{A}\right)\le_{a.s.}\left(X_1^{B},\ldots,X_M^{B}\right),
	$$
	which implies that
	 $$
	\left(X_1^{A},\ldots,X_M^{A}\right)\le_{st}\left(X_1^{B},\ldots,X_M^{B}\right),
	$$
	where $X_{i}^{A(B)}$ represents the number of type $i$ compromised nodes in network $A(B)$, $i=1,\ldots,M$. Therefore, this is proved.
	
	b) Similar to the proof of a), we use $I_{lj,A}=1$ to represent the node $j$, which is compromised by node $l$ in network $A$. Then,
	$$\p(I_{lj,A}=1)=q_{lj}.$$
	Then, we construct   a new network $C$ with the same types of nodes, a propagation depth $L$, and direct compromise probability vector $\bm p$. We assume that the new network $C$ has a different indirect compromise probability matrix $Q'\ge Q$. The matrix $Q'$ is constructed as follows. Let $I_{lj,C}$ represent a node $j$, which is successfully compromised by an indirect attack from node $j$ in network $C$. We define  that if $I_{lj,A}=1$, then $I_{lj,C}=1$; if $I_{lj,A}=0$, then $I_{lj,C}=1$, with probability ${(q_{lj}'-q_{lj})/}(1-q_{lj})$. Then,
	$$\p(I_{lj,C}=1)=q'_{lj}.$$
	Therefore,
	$$
	\left(X_1^{A},\ldots,X_M^{A}\right)\le_{a.s.}\left(X_1^{C},\ldots,X_M^{C}\right),
	$$
	which implies that
	 $$
	\left(X_1^{A},\ldots,X_M^{A}\right)\le_{\rm st}\left(X_1^{C},\ldots,X_M^{C}\right).
	$$
	Thus, the required proof is obtained. \qed

{
 Theorem \ref{thm:infection} implies that when the direct compromise probability increases (e.g., the hacker launches attacks towards newly discovered vulnerabilities of a software or an operating system), more nodes over the network will be compromised. Theorem \ref{thm:infection} further indicates that all types of nodes have a joint larger probability of being compromised, namely, for $\bm p\le \bm p'$,
\begin{eqnarray*}
&&\p(X_1> x_1,\ldots,X_M> x_m|L,\bm p,Q)\\
&\le& \p(X_1> x_1,\ldots,X_M> x_m|L,\bm p',Q).
\end{eqnarray*}

Similarly, when the indirect compromise probability increases (e.g., the malware is very contagious or the network defense is very weak), more nodes will be compromised. Particularly, all types of nodes have a joint larger probability of being compromised.

To sum, we rigorously show that both propagation depth and compromise probabilities (direct or indirect) demonstrate significant effects on the joint cyber risk of a network in the sense of the multivariate stochastic order. When the propagation depth is larger, all types of nodes have larger probabilities of being compromised. Either direct or indirect compromise probability can increase the joint cyber risk of a network.}
 \section{Simulation study}\label{sec:simulation}
{ In this section, we perform a simulation study to assess the joint cyber risk of a network system with heterogeneous nodes. There are three main purposes: (i) The first is to validate the theoretical results in Section \ref{sec:param}. (ii) The second is to provide some new insights on the correlation among risks which are measured via popular dependence measures including Pearson correlation (Pearson), Kendall's tau (Kendall), and Spearman's rho (Spearman) \cite{joe2014dependence}.  (iii) The third is to study the joint cyber risk of a large-scale network via the proposed backward elimination simulation when the explicit computing is infeasible. }

{We generate  a scale-free network with $200$ nodes according to Barabasi--Albert model \cite{BA1999}. We first set five initial nodes, and then each new node is randomly connected to two existing nodes with probabilities proportional to the degrees of the existing nodes.} The generated graph is shown in Figure \ref{200ba}.
We choose 20 nodes with the top highest degrees as the type I nodes, and others as the type II nodes. The degrees of the type I nodes are: 49,38,37,35,35,33,32,30,28,24,24,21,19,18,17,16,16,15,15,15.
	
\begin{figure}[hbtp!]
\centering
\includegraphics[width=.4\textwidth] {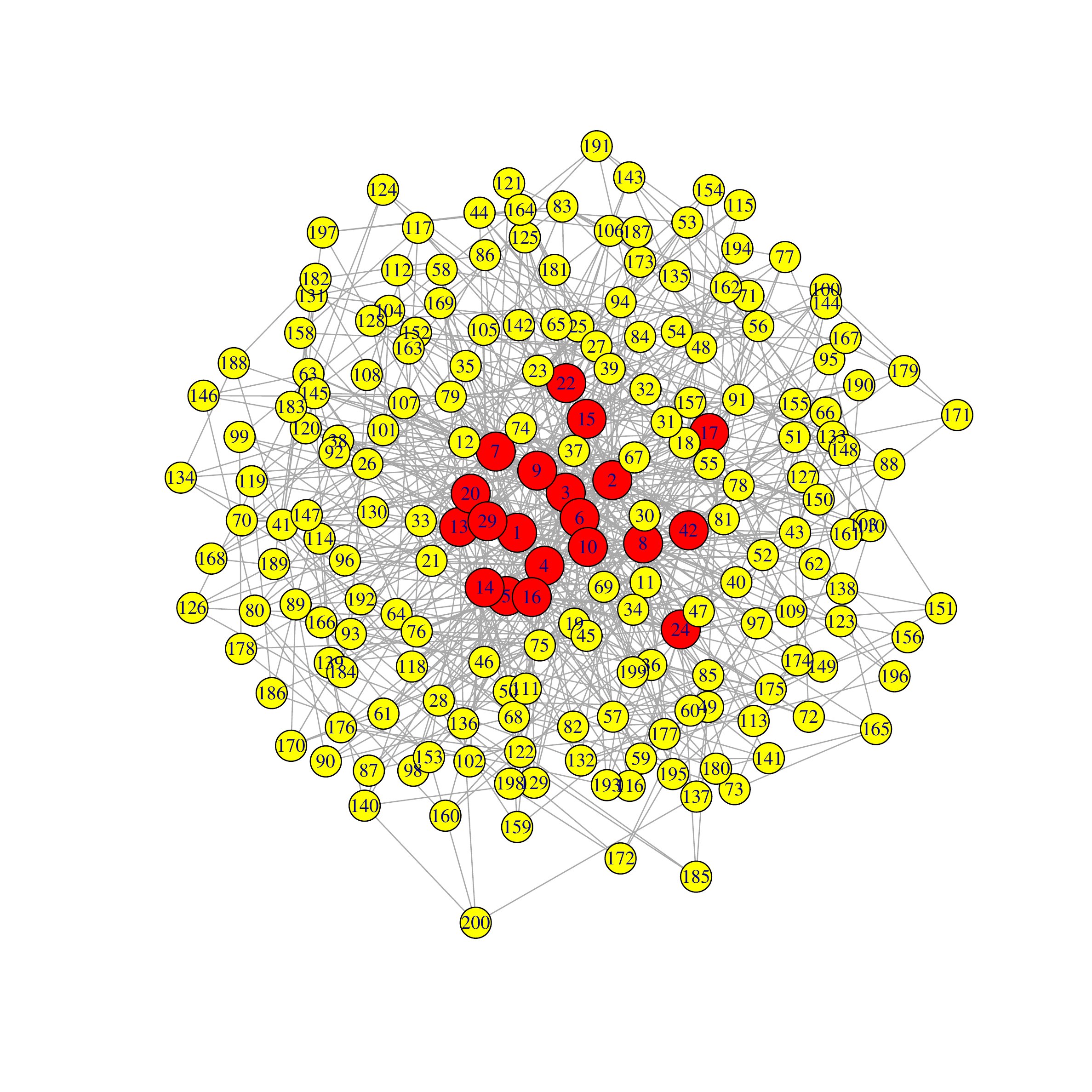}
\caption{Scale-free network with 200 nodes, where the red color nodes represent the type I nodes, and the others are type II nodes \label{200ba}}
\end{figure}

Let $X_i$ denote the varying numbers of compromised nodes of type $i$, $i=1,2$, and $X=X_1+X_2$. For simplicity, we assume  that the direct compromise probabilities are the same for each type, i.e., $p_I$ and $p_{II}$ for type I and type II nodes, respectively.
Similarly, we assume that the indirect compromise probabilities are the same for each type, i.e., $q_I$ and $q_{II}$ for type I and type II nodes, respectively.

 {Algorithm \ref{alg:sim} is used to simulate the number of compromised nodes. The simulation algorithm is primarily based on the proposed backward elimination approach. In Algorithm \ref{alg:sim}, for a given $L$, all the compromised nodes for every depth $1\le l\le L$  are recorded (lines 14 and 16); thus, obtaining all the different results for $l$. The experiment is performed based on $100,000$ Monte Carlo simulations.}
{\color{blue}
	\begin{algorithm}[!htbp]
	\caption{Algorithm for simulating compromised nodes via the back elimination approach.	\label{alg:sim}}
	INPUT: network graph $G(V,E)$; number of nodes $N$; propagation depth $L$; directly compromise probability vector ${\bm p}$; propagation probability matrix $Q$; number of iterations $K$\\
	OUTPUT: Simulated number of compromised nodes of all types
	{
	\begin{algorithmic} [1]
	\FOR{$k\in 1:K$}
	\STATE{Determine the nodes that are directly compromised, i.e., generate Bernoulli random vector $(I_1,\ldots,I_N)$ with probability ${\bm p}$, and let $U_0=\{i:I_i=1\}$ denote the set of compromised nodes;}
	\FOR{$h\in 1:L$}
	\FOR{$j\in V\backslash U_0$}
	\STATE{Determine whether node $j$ is compromised by the nodes in $U_0$ with propagation probability matrix $Q$, i.e., generate independent Bernoulli random variables $I_{l,j}$ with probabilities $q_{lj}$, $l\in U_0$;}
	\IF{Node $j$ is compromised, i.e., $\max_{l\in U_0}\{I_{lj}\}=1$;}
	\STATE{Record the new compromised nodes into set $U_{h}$ ;}
	\ENDIF
	\ENDFOR
	\STATE{Count the number of compromised nodes for each type in $U_h$, denoted by
	$\left(u^{(k)}_{h,1},\ldots,u^{(k)}_{h,M}\right)$;}
	\STATE{Update $V\leftarrow V\backslash U_0$, i.e., by removing the nodes $U_0$ from $V$, and all the edges connecting $U_0$;}
	\STATE{Update $U_0\leftarrow U_{h}$;}
		\ENDFOR
	\STATE{Compute the number of compromised nodes for each round and type as
	$\left(x^{(k)}_{l,1},\ldots,x^{(k)}_{l,M}\right)=\left(\sum_{h=0}^lu^{(k)}_{h,1},\ldots,\sum_{h=0}^lu^{(k)}_{h,M}\right)$, $l=1,\ldots,L$;}
	\ENDFOR
	\RETURN{$\left(x^{(k)}_{l,1},\ldots,x^{(k)}_{l,M}\right)$, $k=1,\ldots,K$, $l=1,\ldots,L$.}
	\end{algorithmic}
	}
	\end{algorithm}
	}

\begin{figure*}[!htbp]
\centering
\subfigure[Mean, $\zeta=(0.05,0.15,0.2,0.3)$]{\includegraphics[width=0.31\textwidth]{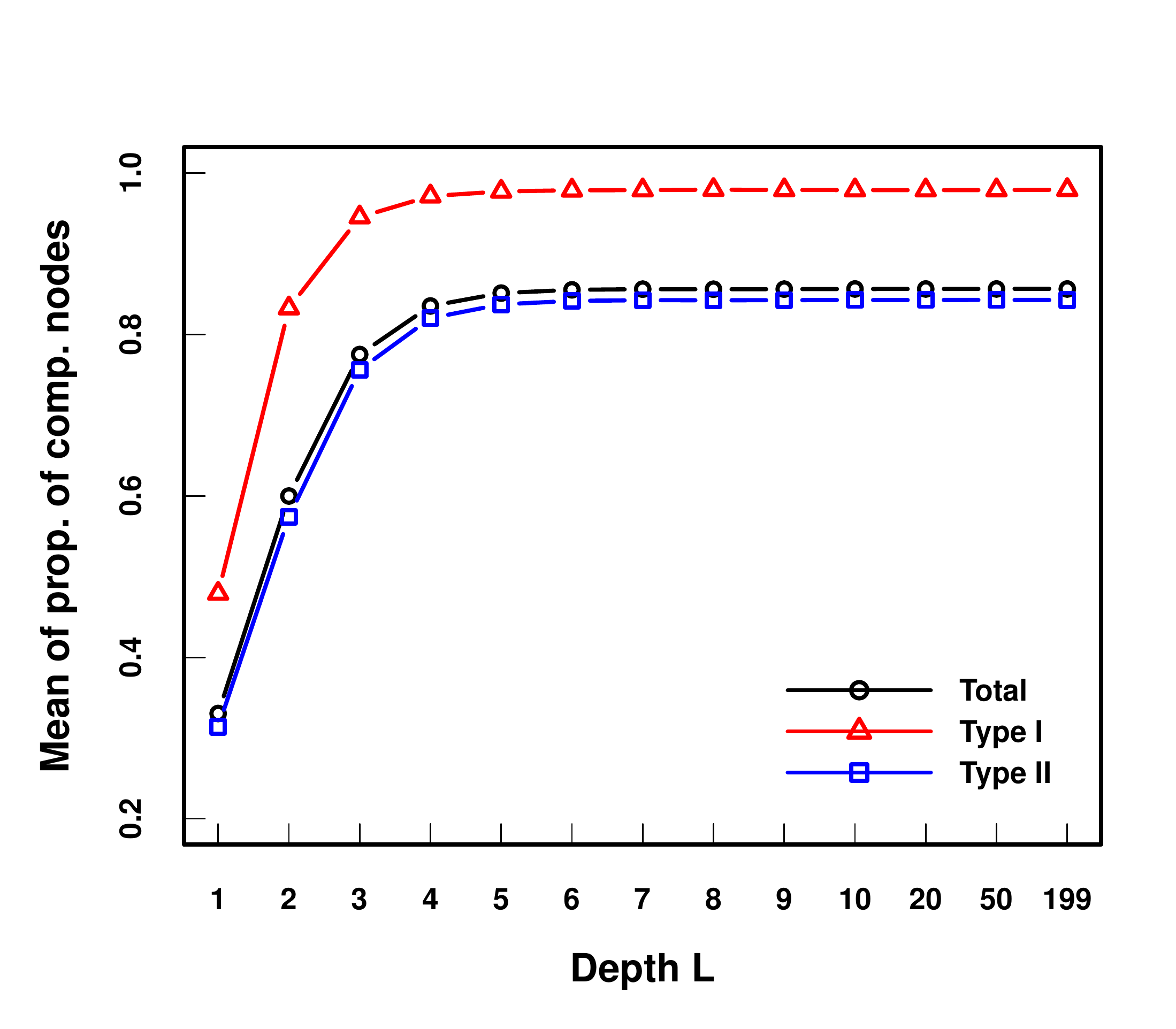}\label{fig:p1mean}}\quad
\subfigure[SD, $\zeta=(0.05,0.15,0.2,0.3)$]{\includegraphics[width=0.31\textwidth]{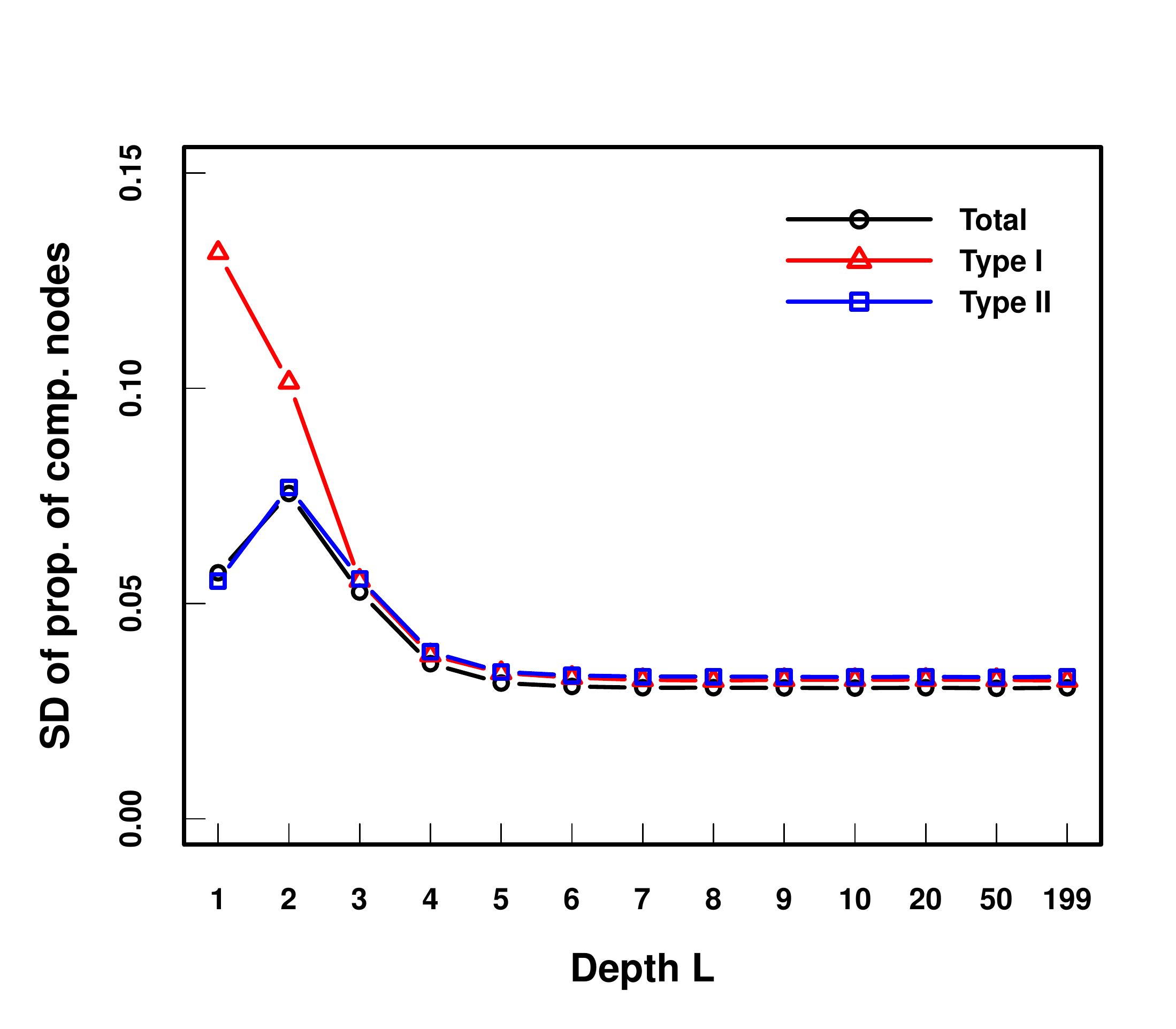}\label{fig:p1sd}}\quad
\subfigure[Corr., $\zeta=(0.05,0.15,0.2,0.3)$]{\includegraphics[width=0.31\textwidth]{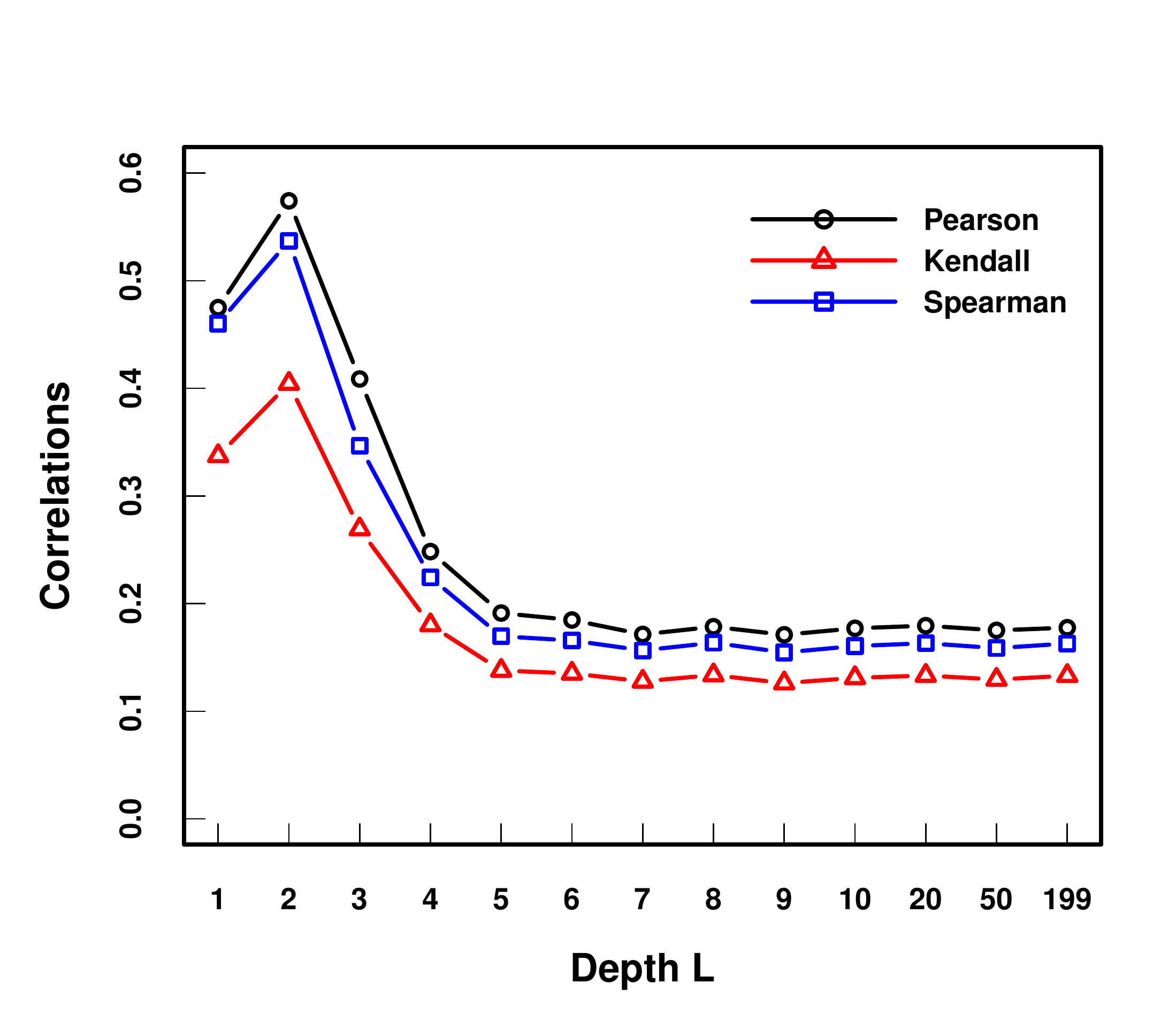}\label{fig:p1cor}}\quad
\subfigure[Mean, $\zeta=(0.05,0.15,0.4,0.5)$]{\includegraphics[width=0.31\textwidth]{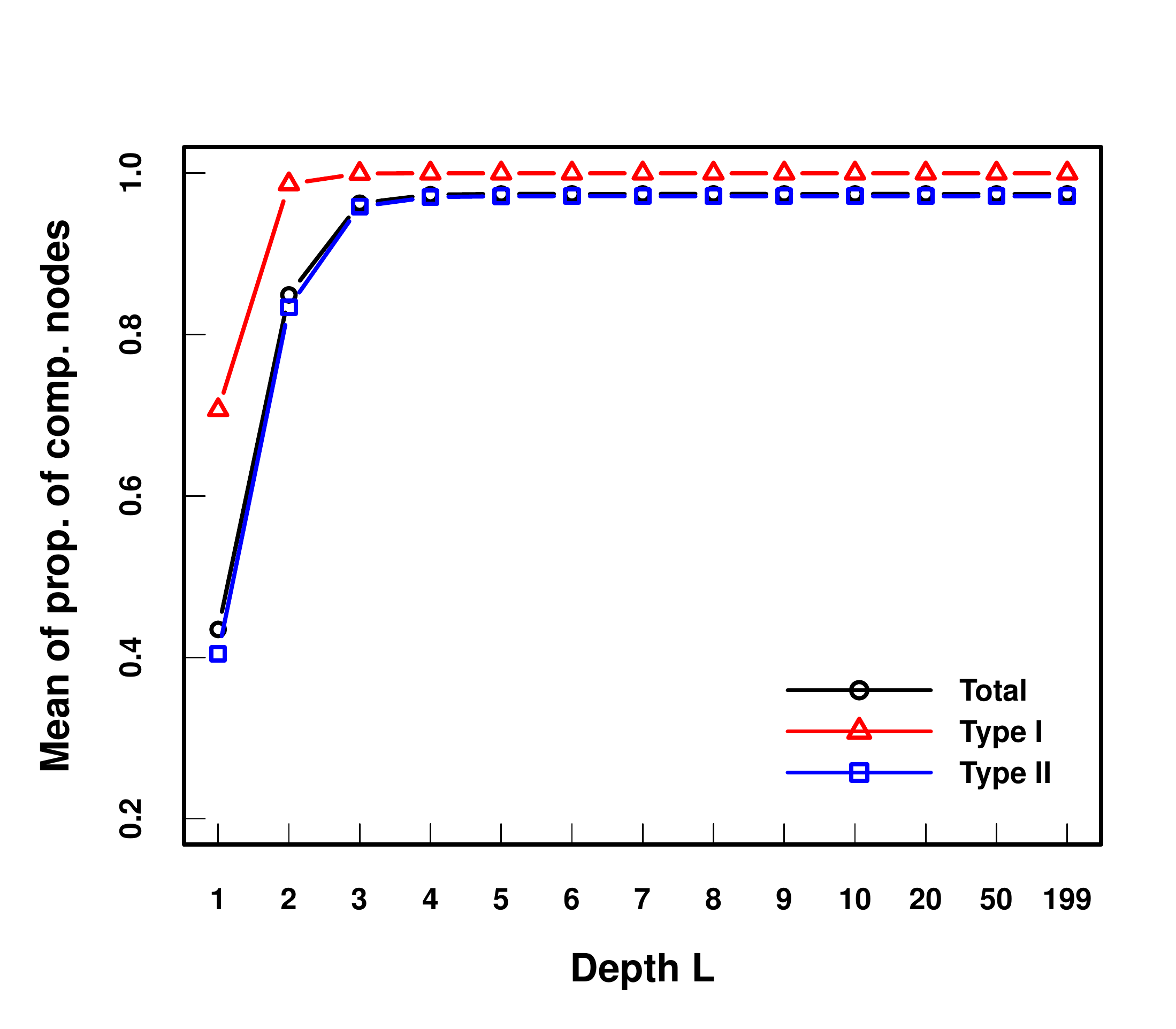}\label{fig:p2mean}}\quad
\subfigure[SD, $\zeta=(0.05,0.15,0.4,0.5)$]{\includegraphics[width=0.31\textwidth]{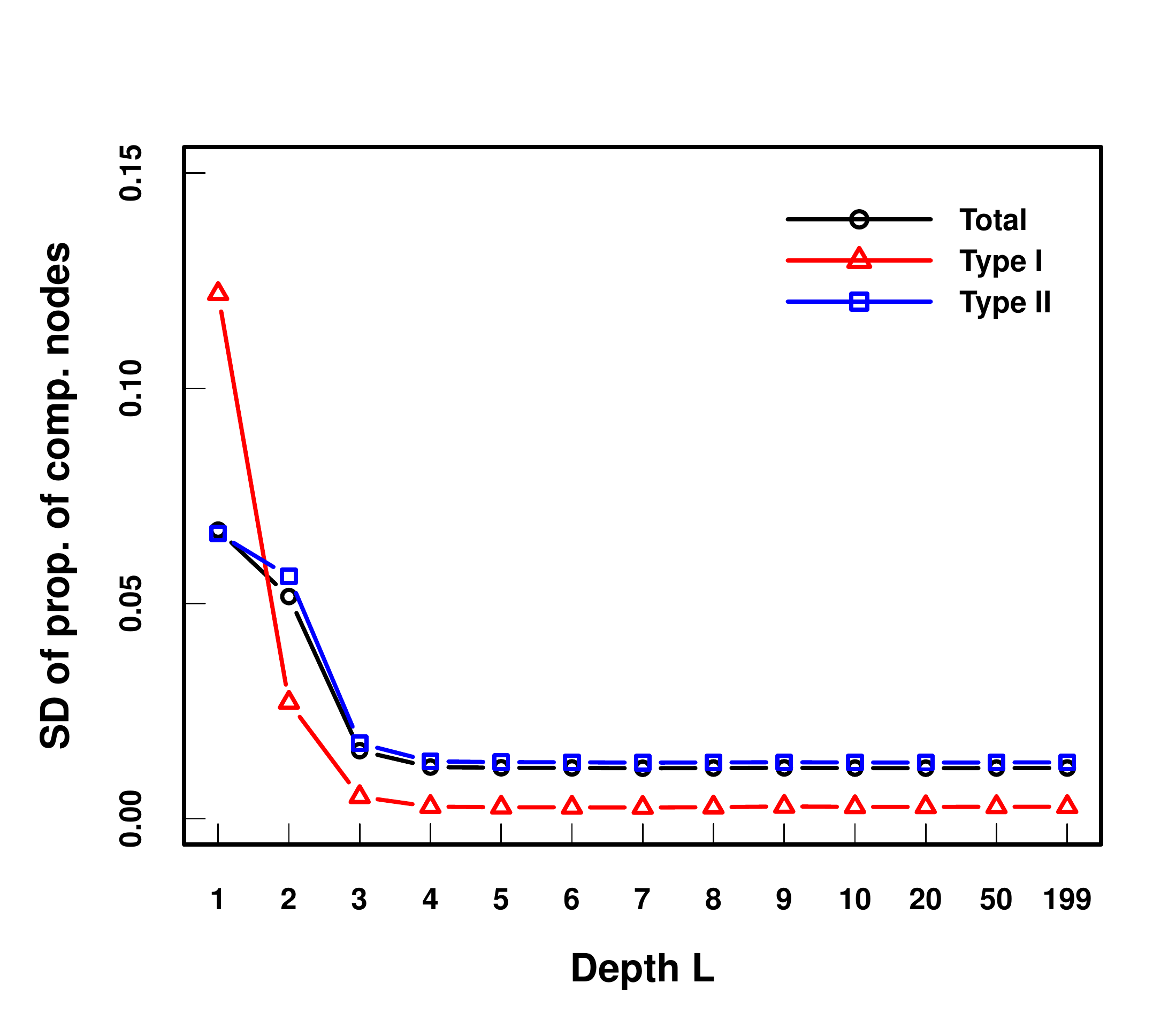}\label{fig:p2sd}}\quad
\subfigure[Corr., $\zeta=(0.05,0.15,0.4,0.5)$]{\includegraphics[width=0.31\textwidth]{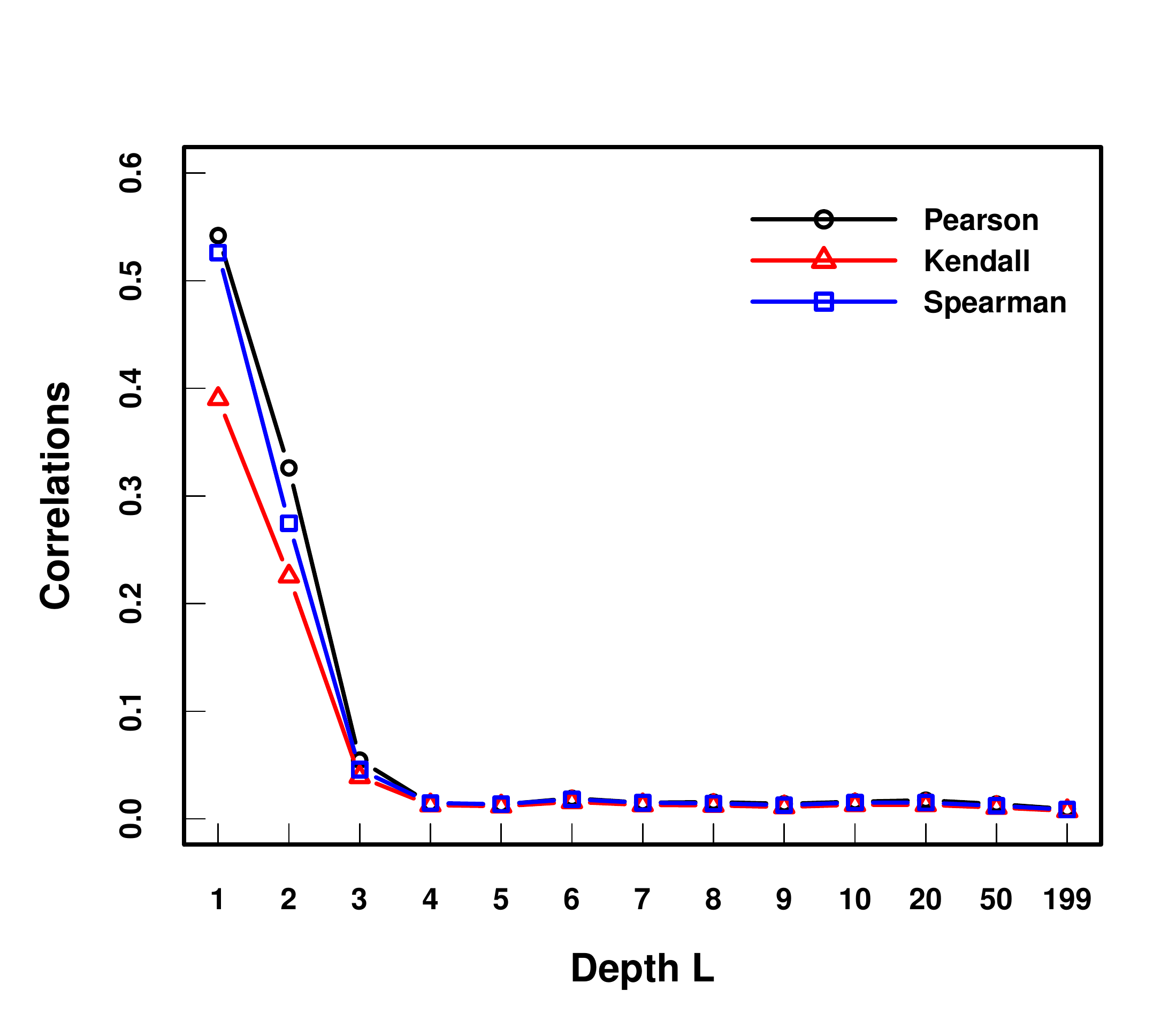}\label{fig:p2cor}}\quad
\subfigure[Mean, $\zeta=(0.1,0.2,0.2,0.3)$]{\includegraphics[width=0.31\textwidth]{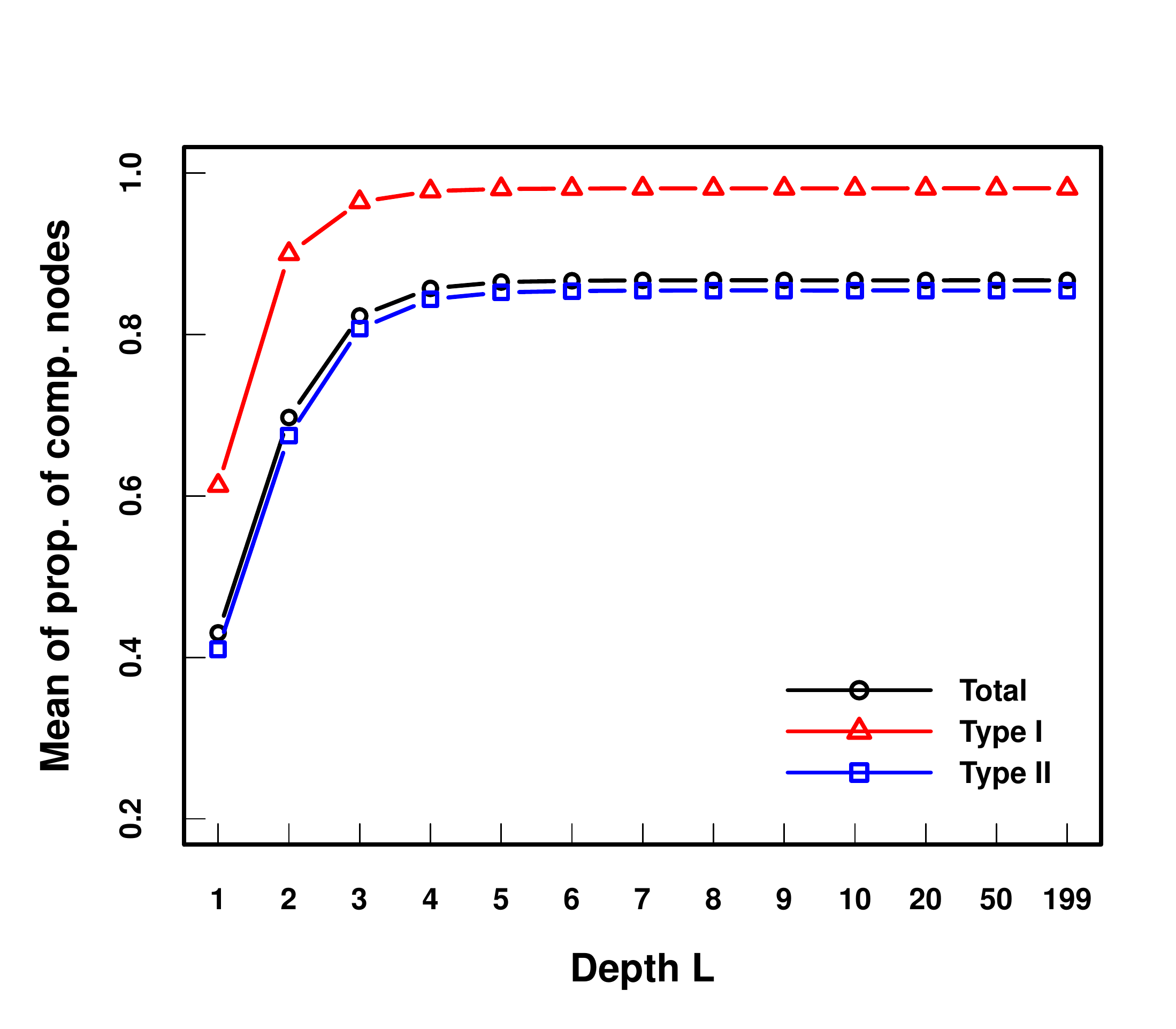}\label{fig:p3mean}}\quad
\subfigure[SD, $\zeta=(0.1,0.2,0.2,0.3)$]{\includegraphics[width=0.31\textwidth]{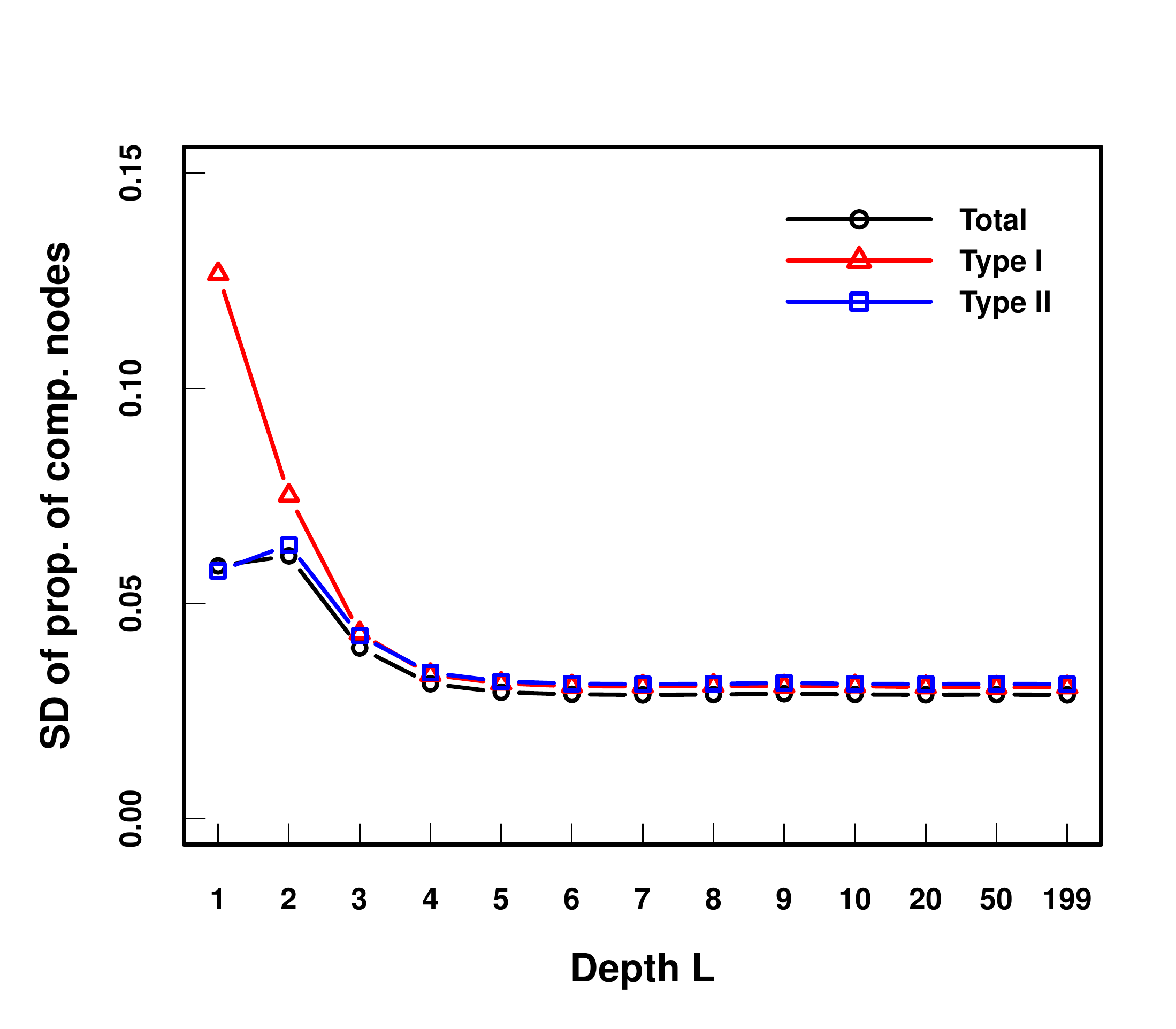}\label{fig:p3sd}}\quad
\subfigure[Corr., $\zeta=(0.1,0.2,0.2,0.3)$]{\includegraphics[width=0.31\textwidth]{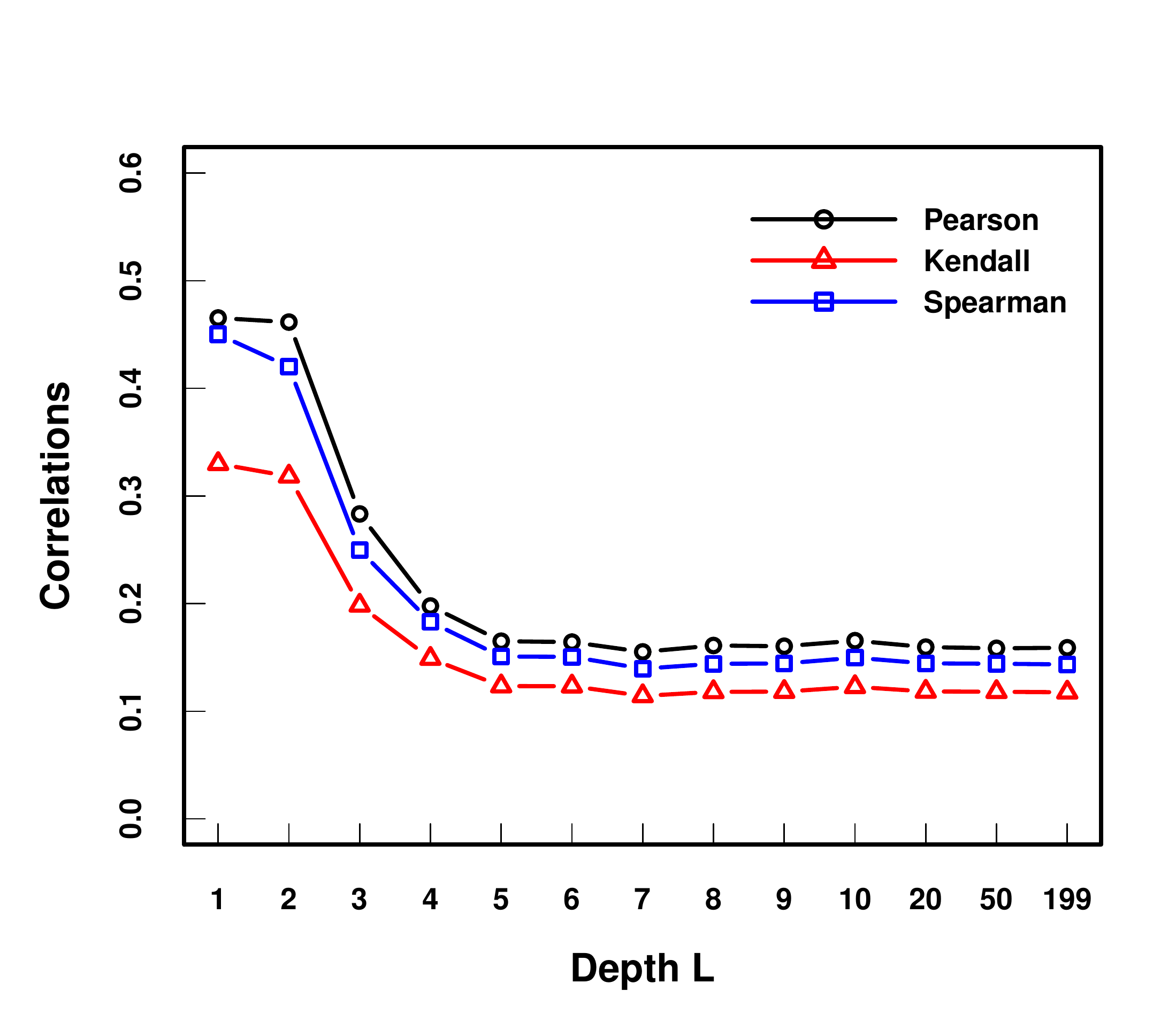}\label{fig:p3cor}}
\caption{Means, standard deviations (SD), and correlations (Corr.) of proportions of compromised nodes for different parameters
$\zeta= (p_I,p_{II},q_I,q_{II})$. \label{fig:pmf1}}
\end{figure*}

{
\subsection{Propagation depth}  Figure \ref{fig:pmf1} shows the means, standard deviations, and correlations of proportions of compromised nodes under different scenarios.  It is seen from Figures  \ref{fig:p1mean},  \ref{fig:p2mean}, and \ref{fig:p3mean} that the means of proportions of compromised nodes increase with the propagation depth $L$ for both type I and type II nodes. In particular, we observe that the means of proportions of compromised nodes increases dramatically from $L=1$ to $L=2$ (around $25\%$ to $40\%$).  However, when $L$ increases from $4$ to $20$ or more, the mean of  proportions of compromised nodes only varies marginally. This is primarily because the dynamic of compromise has attained a relatively steady state. For type I nodes, almost all nodes are compromised, while for type II nodes, there are still a few healthy nodes. This suggests that the propagation depth $L$ has a significant effect on the number of compromised nodes.

While considering the standard deviation from Figures  \ref{fig:p1sd},  \ref{fig:p2sd}, and \ref{fig:p3sd}, it is observed that for a smaller $L$ (namely, 1 and 2), the standard deviations of the proportions of compromised nodes are larger for both types. This is because, when $L$
is small, the varying numbers of compromised nodes could be quite different. However, when $L$ is larger (say, greater or equal to $4$), the standard deviation is smaller, owing to the fact that the dynamic of compromise has become a relatively steady state, i.e., the proportions of compromised nodes are nearly the same. We conclude  that the propagation depth $L$ has a significant effect on the standard deviations of  proportions of compromised nodes.

For the correlations between the proportions of compromised nodes of  both types, it is observed from Figures  \ref{fig:p1cor},  \ref{fig:p2cor}, and \ref{fig:p3cor}  that the correlation is larger when $L$ is small as  measured by the three dependence measures.   When $L$ is larger, the correlation becomes smaller. This observations can be explained as follows: when $L$ is smaller, a larger  proportion of compromised type I nodes should be associated with a larger proportion of compromised type II nodes; when $L$ is large, the  proportions of compromised nodes for both types are relatively stable, which results in a smaller correlation.

\subsection{Compromise probabilities} From Figure \ref{fig:pmf}, it is seen that all the probability curves are unimodal and they shift to the right when $L$ is large. In particular, we observe that the probability curves have the largest peak values for $L=10$.  Comparing the Figures \ref{fig:p1t1}, \ref{fig:p1t2} and \ref{fig:p1tot} to those corresponding ones in the second row of Figure \ref{fig:pmf}, it is observed that the larger indirect compromise probabilities cause the probability curves shifting to the right. This implies that when the indirect compromise probabilities increase, more nodes are compromised.  Comparing the Figures \ref{fig:p1t1}, \ref{fig:p1t2} and \ref{fig:p1tot} to those corresponding ones in the third row of Figure \ref{fig:pmf}, the similar conclusion can be drawn for the direct compromise probabilities.
These observations validate the theoretical results of Theorem \ref{thm:infection}.

\begin{figure*}[!htbp]
\centering
\subfigure[Type I, $\zeta=(0.05,0.15,0.2,0.3)$]{\includegraphics[width=0.31\textwidth]
{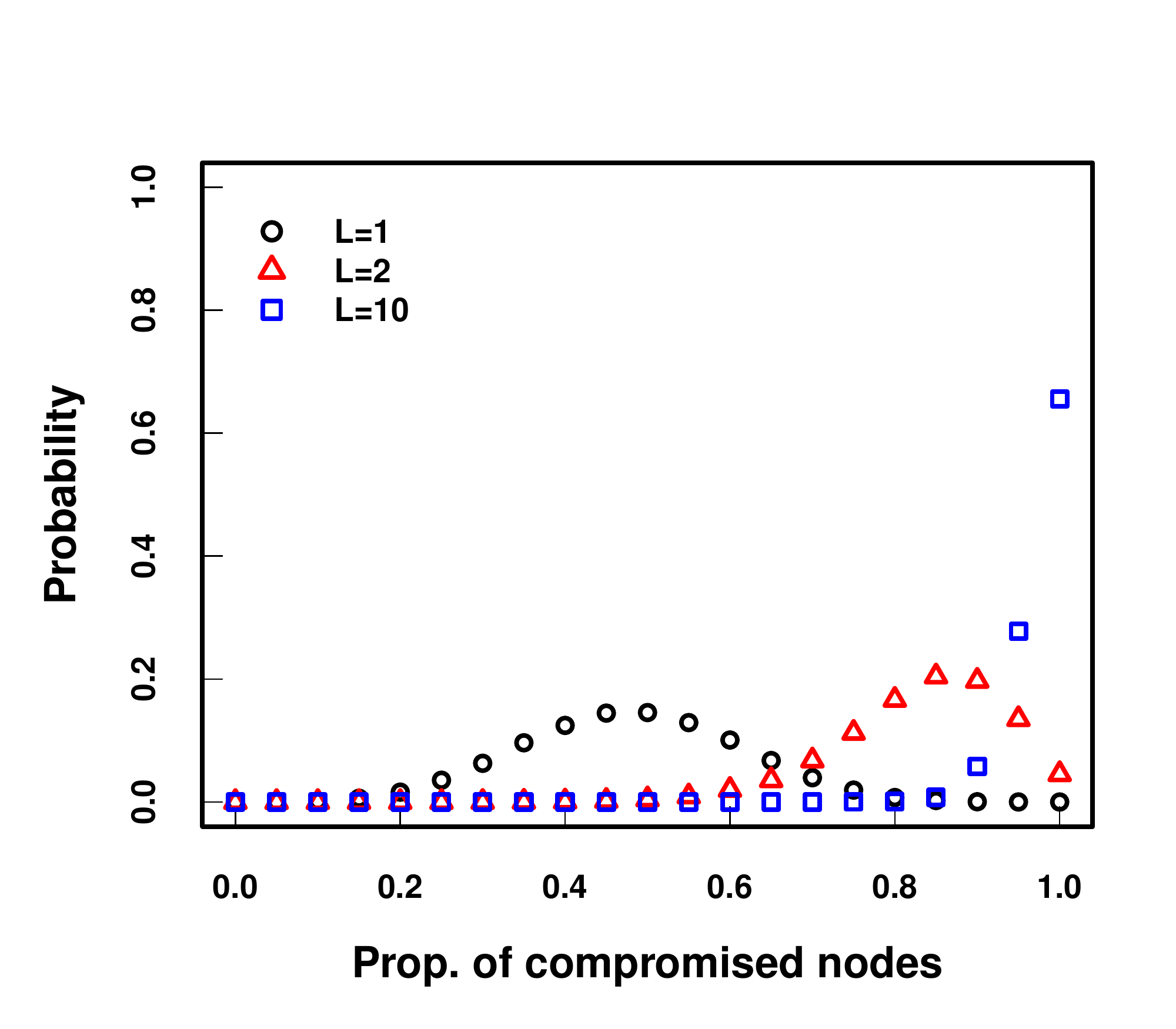}\label{fig:p1t1}}\quad
\subfigure[Type II, $\zeta=(0.05,0.15,0.2,0.3)$]{\includegraphics[width=0.31\textwidth]{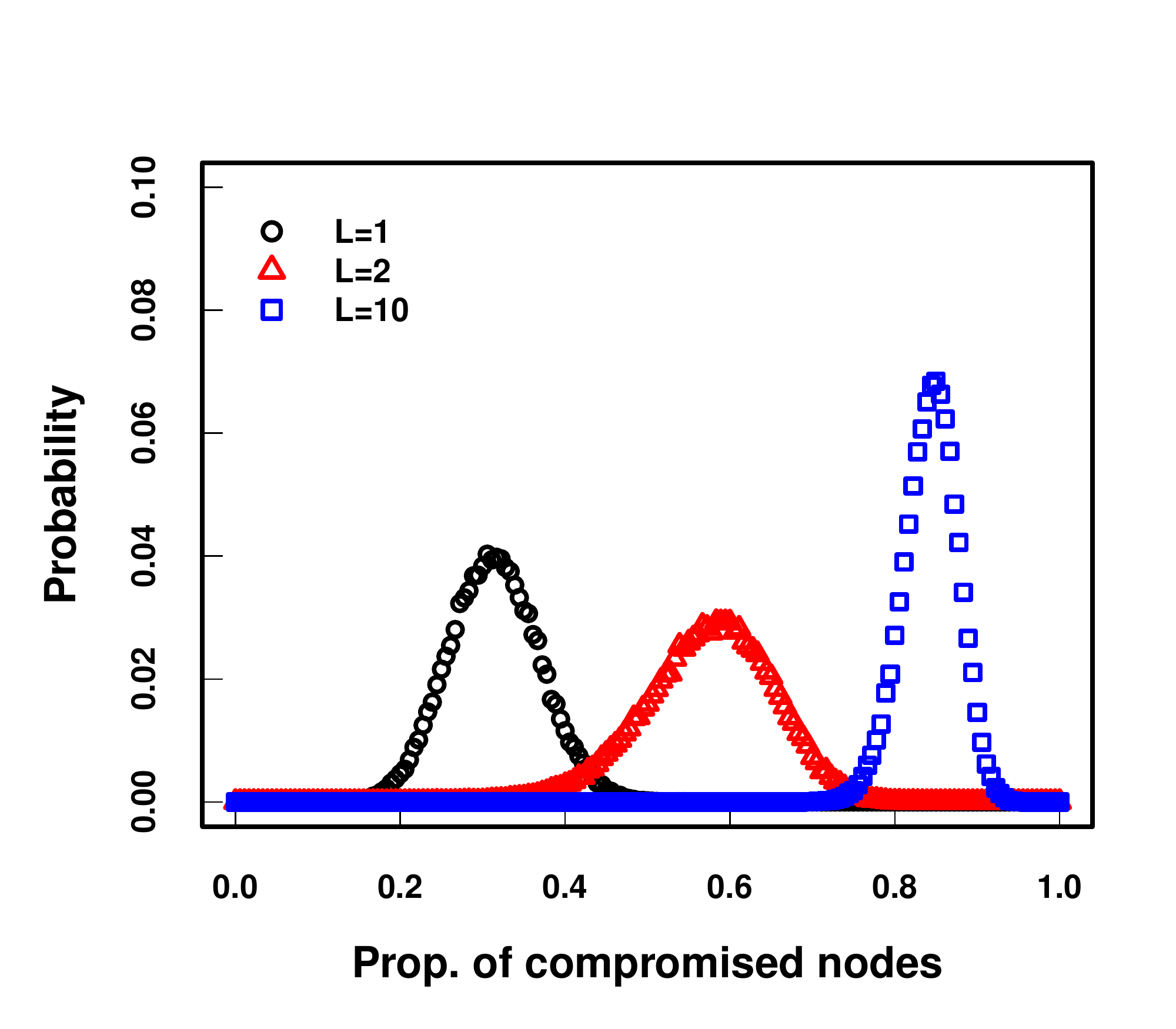}
\label{fig:p1t2}}\quad
\subfigure[Total, $\zeta=(0.05,0.15,0.2,0.3)$]{\includegraphics[width=0.31\textwidth]
{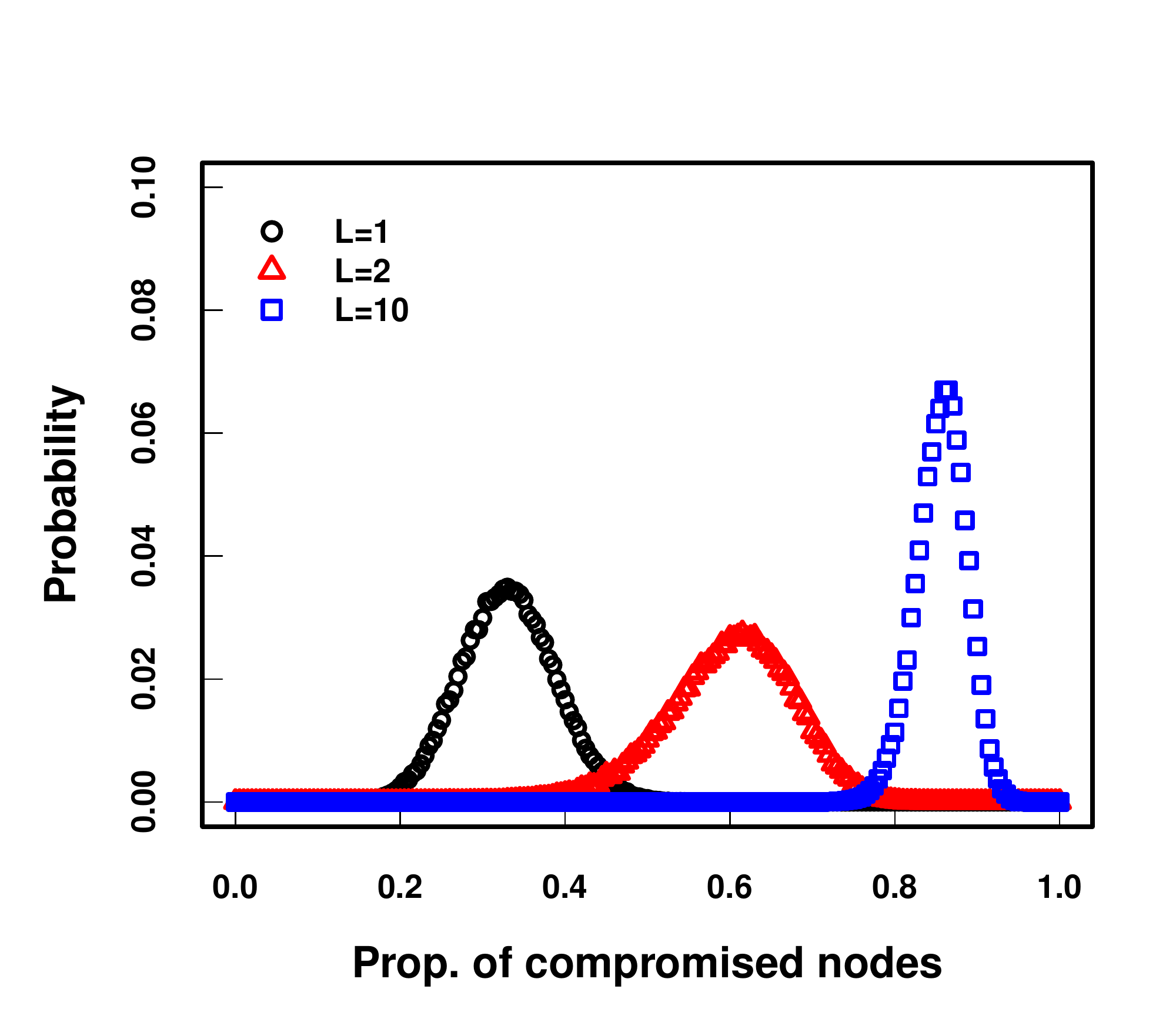}\label{fig:p1tot}}\quad
\subfigure[Type I, $\zeta=(0.05,0.15,0.4,0.5)$]{\includegraphics[width=0.31\textwidth]
{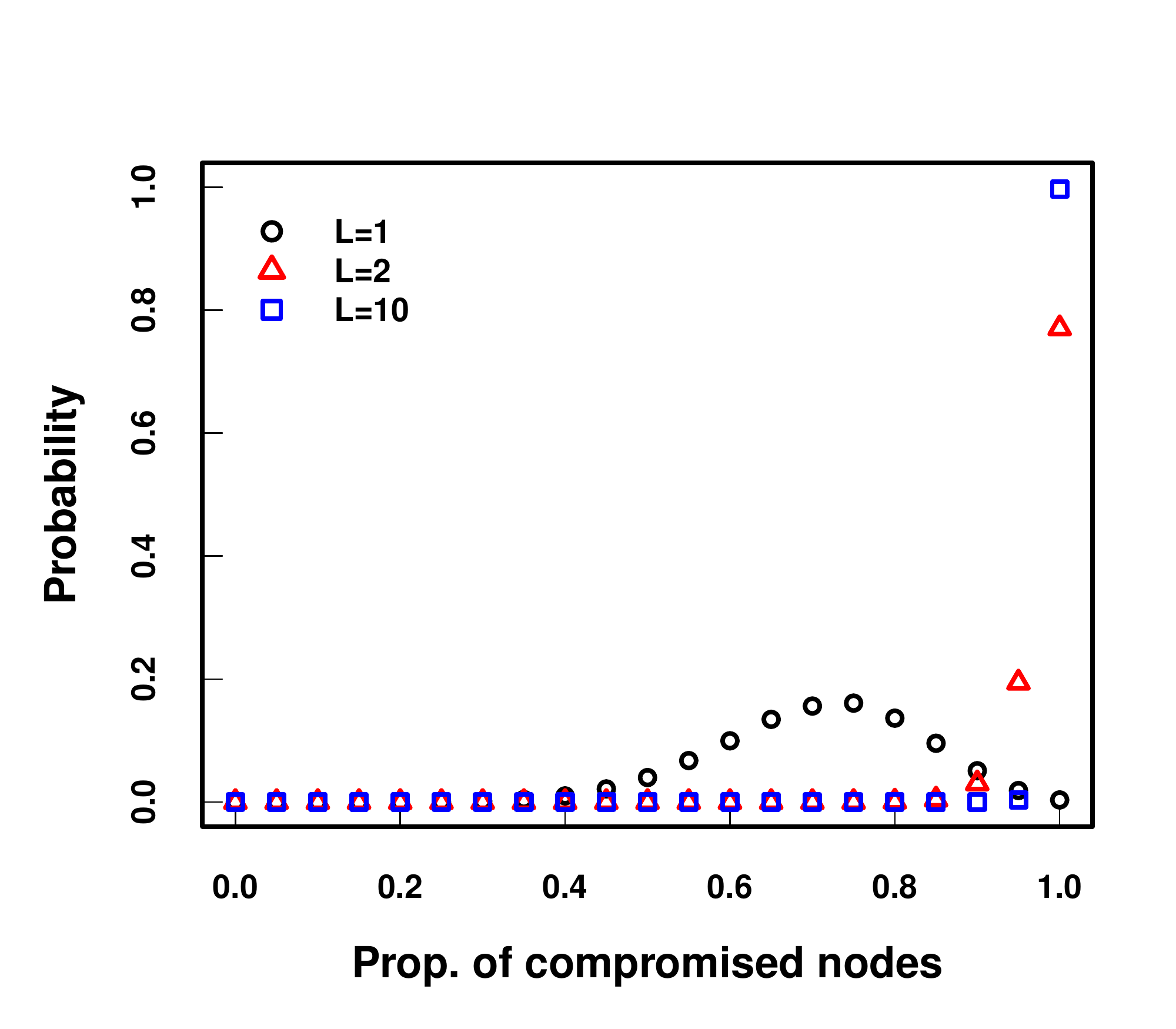}\label{fig:p2t1}}\quad
\subfigure[Type II, $\zeta=(0.05,0.15,0.4,0.5)$]{\includegraphics[width=0.31\textwidth]
{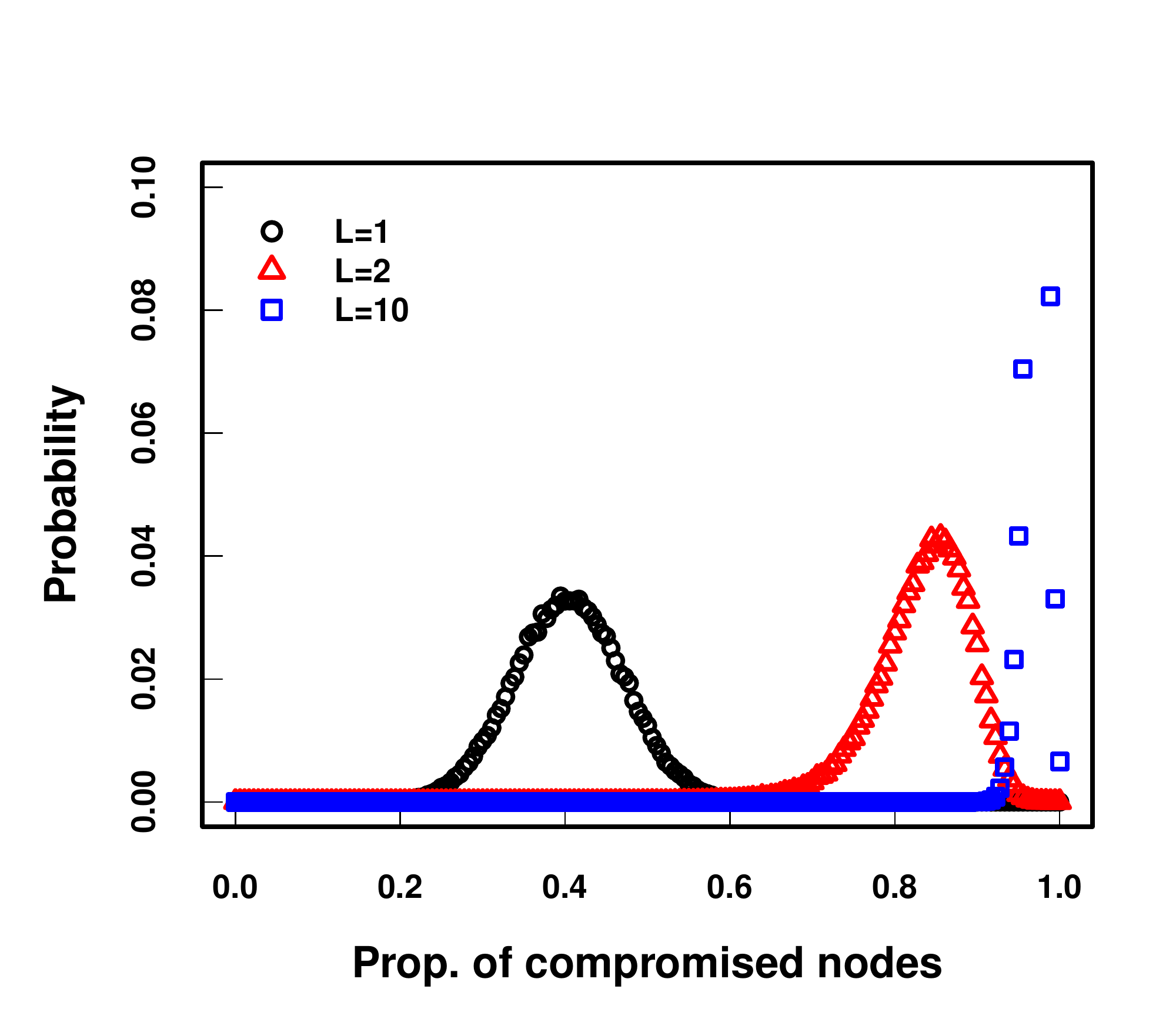}\label{fig:p2t2}}\quad
\subfigure[Total, $\zeta=(0.05,0.15,0.4,0.5)$]{\includegraphics[width=0.31\textwidth]
{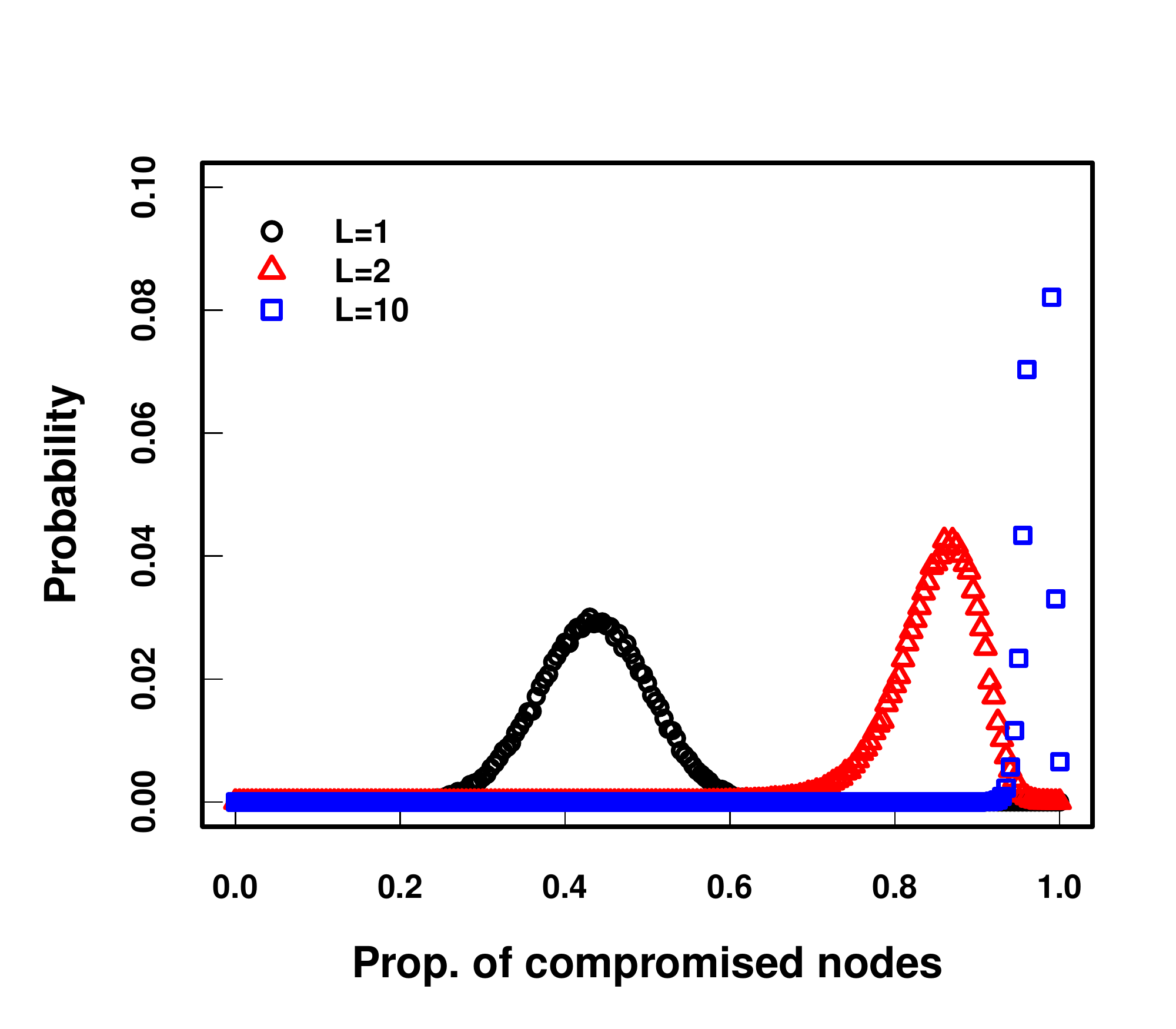}\label{fig:p2tot}}\quad
\subfigure[Type I, $\zeta=(0.1,0.2,0.2,0.3)$]{\includegraphics[width=0.31\textwidth]
{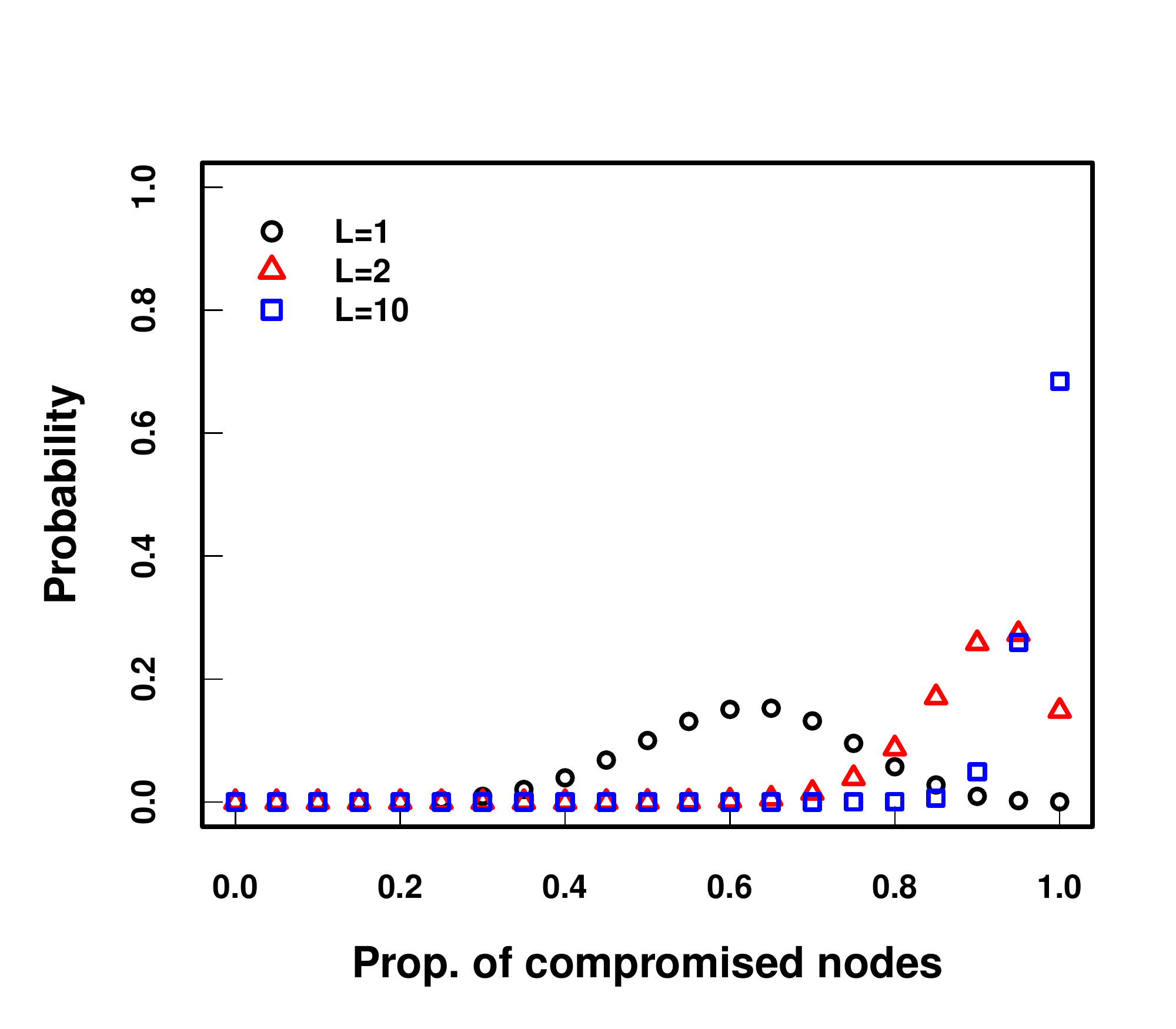}\label{fig:p3t1}}\quad
\subfigure[Type II, $\zeta=(0.1,0.2,0.2,0.3)$]{\includegraphics[width=0.31\textwidth]
{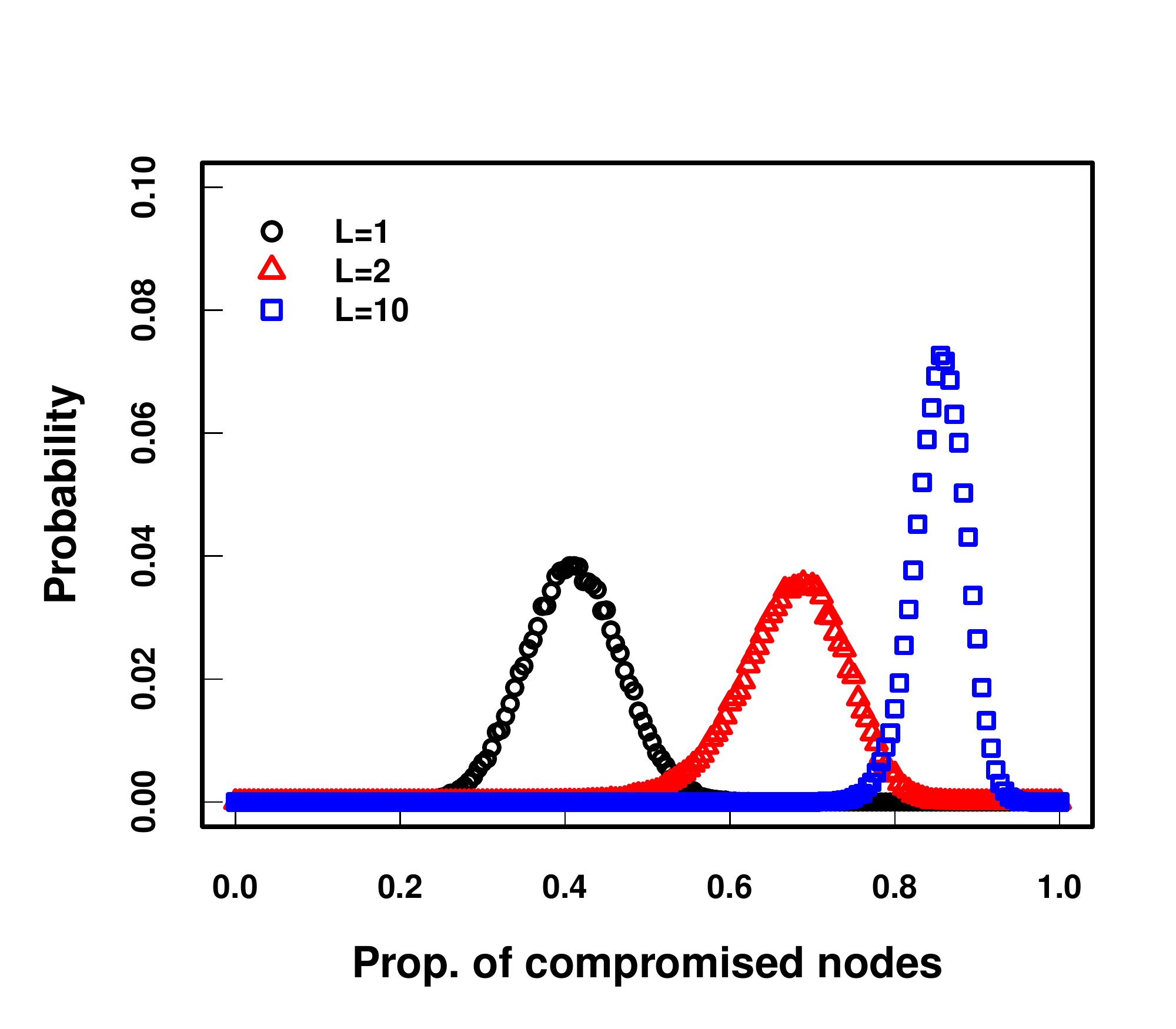}\label{fig:p3t2}}\quad
\subfigure[Total, $\zeta=(0.1,0.2,0.2,0.3)$]{\includegraphics[width=0.31\textwidth]{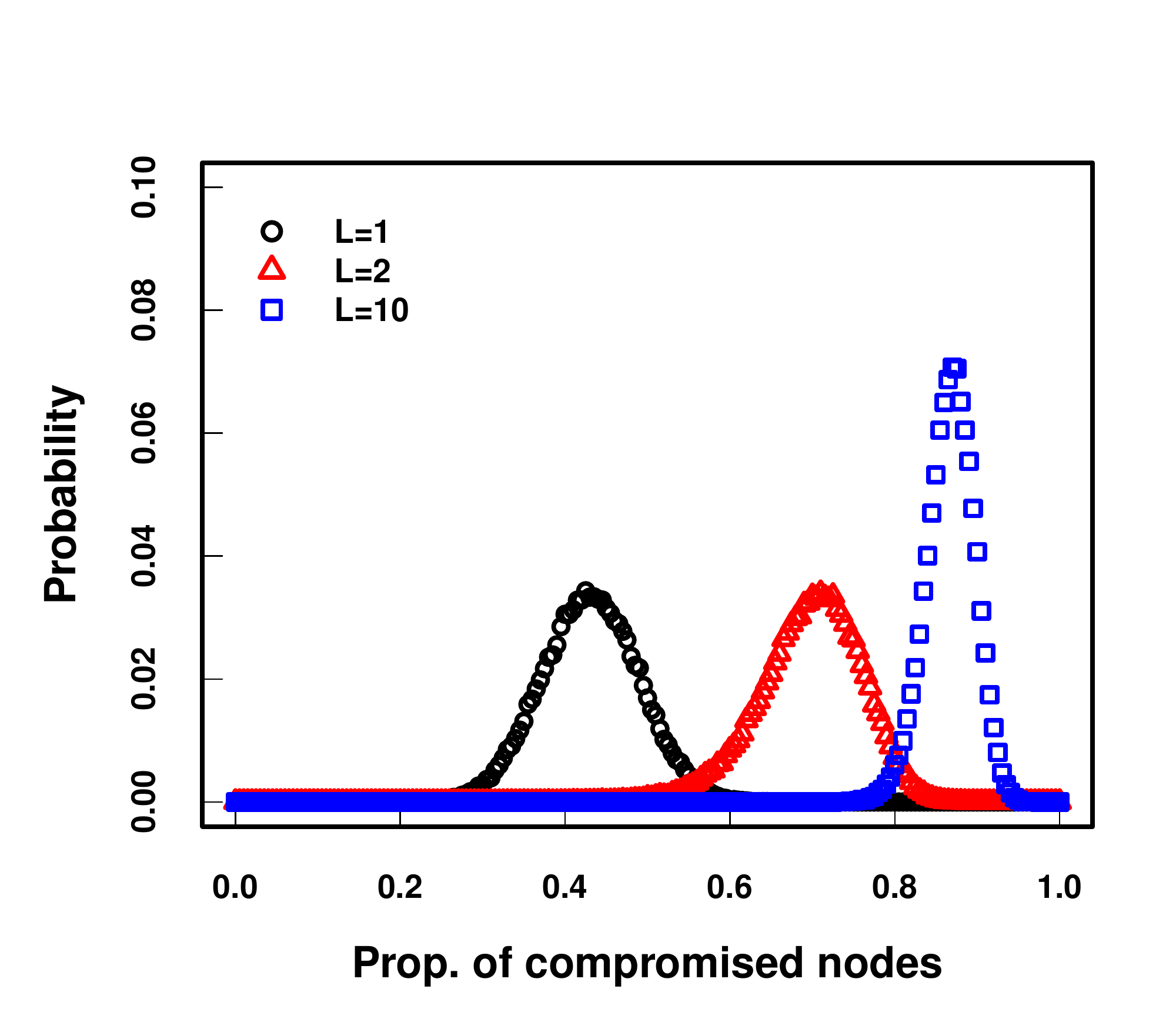}\label{fig:p3tot}}
\caption{Probability plots of proportions of compromised nodes for different $\zeta= (p_I,p_{II},q_I,q_{II})$ and $L=1,2,10$\label{fig:pmf}}
\end{figure*}

\subsection{Joint cyber risks} To assess the joint cyber risks of type I and type II nodes, we display the contour plots of the joint probabilities of compromised nodes in Figure \ref{fig:contour1} with  $$(p_I,p_{II},q_I,q_{II})=(0.05,0.15,0.2,0.3).$$

Figure \ref{fig:joint1} shows the contour plot for the case of $L=1$. It is observed that the proportions of compromised nodes of different types demonstrate a clear positive correlation. Thus, when the proportion of type I compromised nodes increases, the proportion of type II compromised nodes also increases. By comparing Figure \ref{fig:joint1} and Figure \ref{fig:joint2}, we observe that there also exists a positive dependence between the compromised proportions. It is very clear that the joint risk  significantly increases when $L=2$. Figure \ref{fig:joint10} shows the contour plot for the case of $L=10$. It is observed that the joint risk is the highest among all the cases. However, it does not show any clear positive dependence pattern between two types of nodes, which coincides with the previous correlation analysis.
\begin{figure*}[!htbp]
\centering
 \subfigure[$L=1$]{\includegraphics[width=0.31\textwidth]{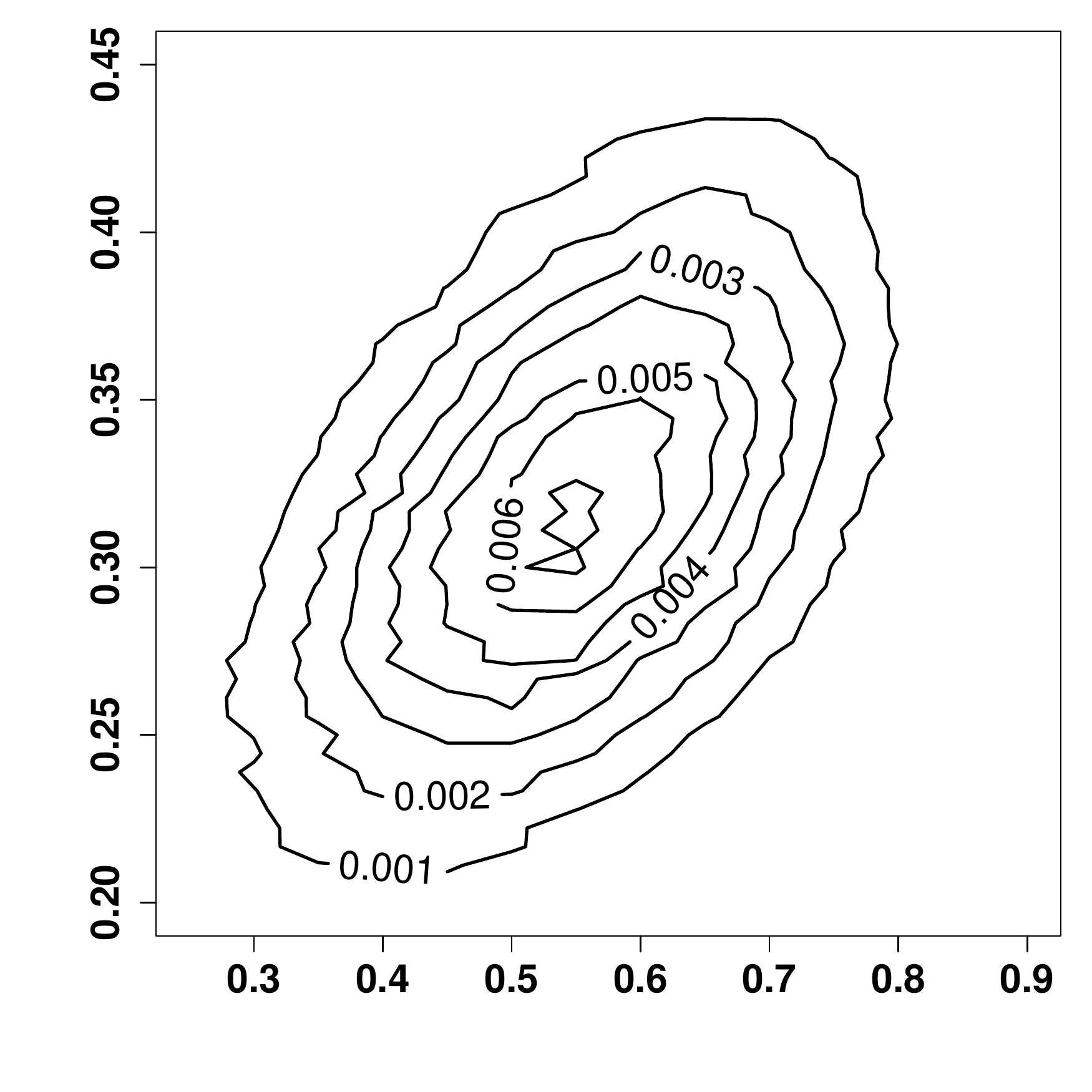}\label{fig:joint1}}
 \subfigure[$L=2$]{\includegraphics[width=0.31\textwidth]{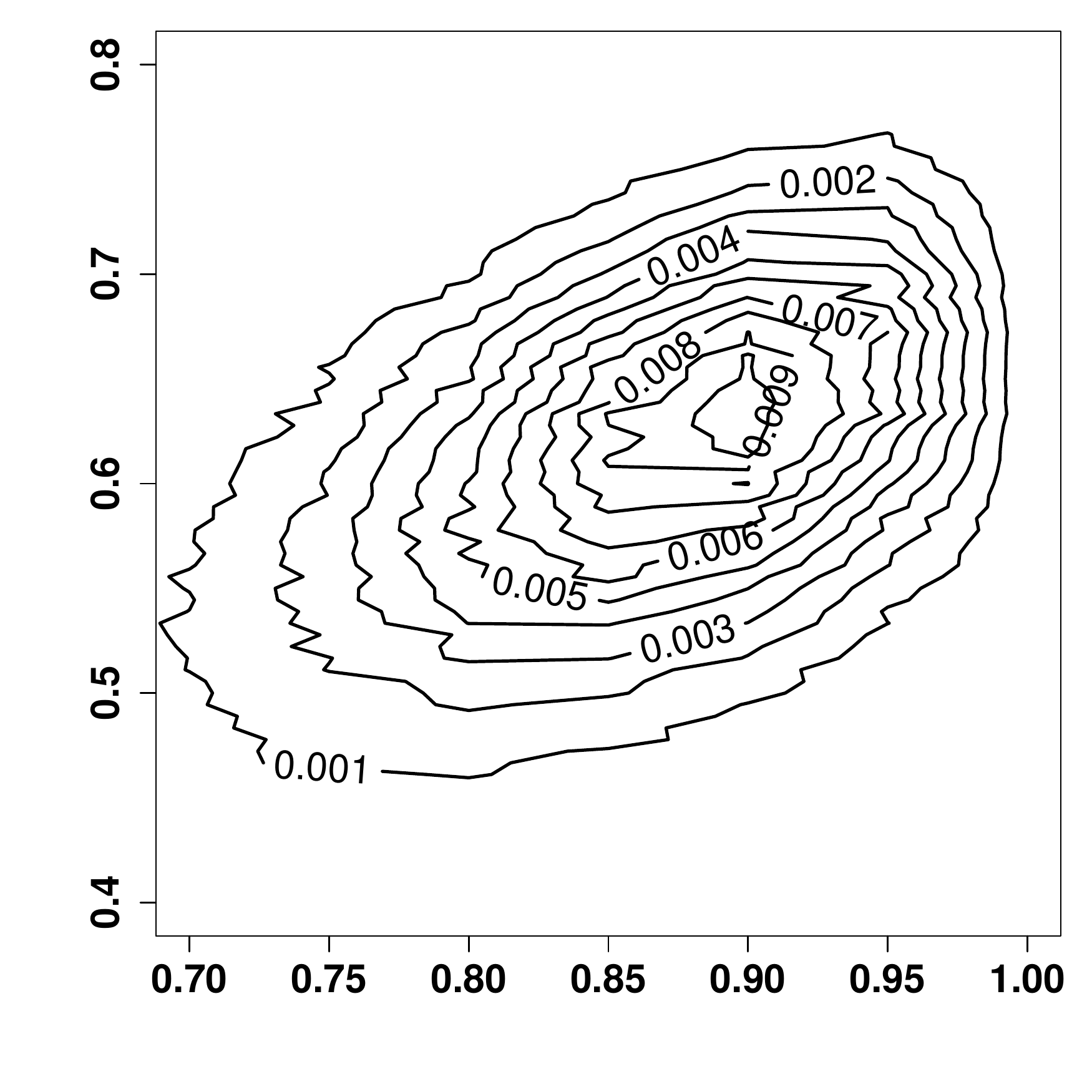}\label{fig:joint2}}
 \subfigure[$L=10$]{\includegraphics[width=0.31\textwidth]{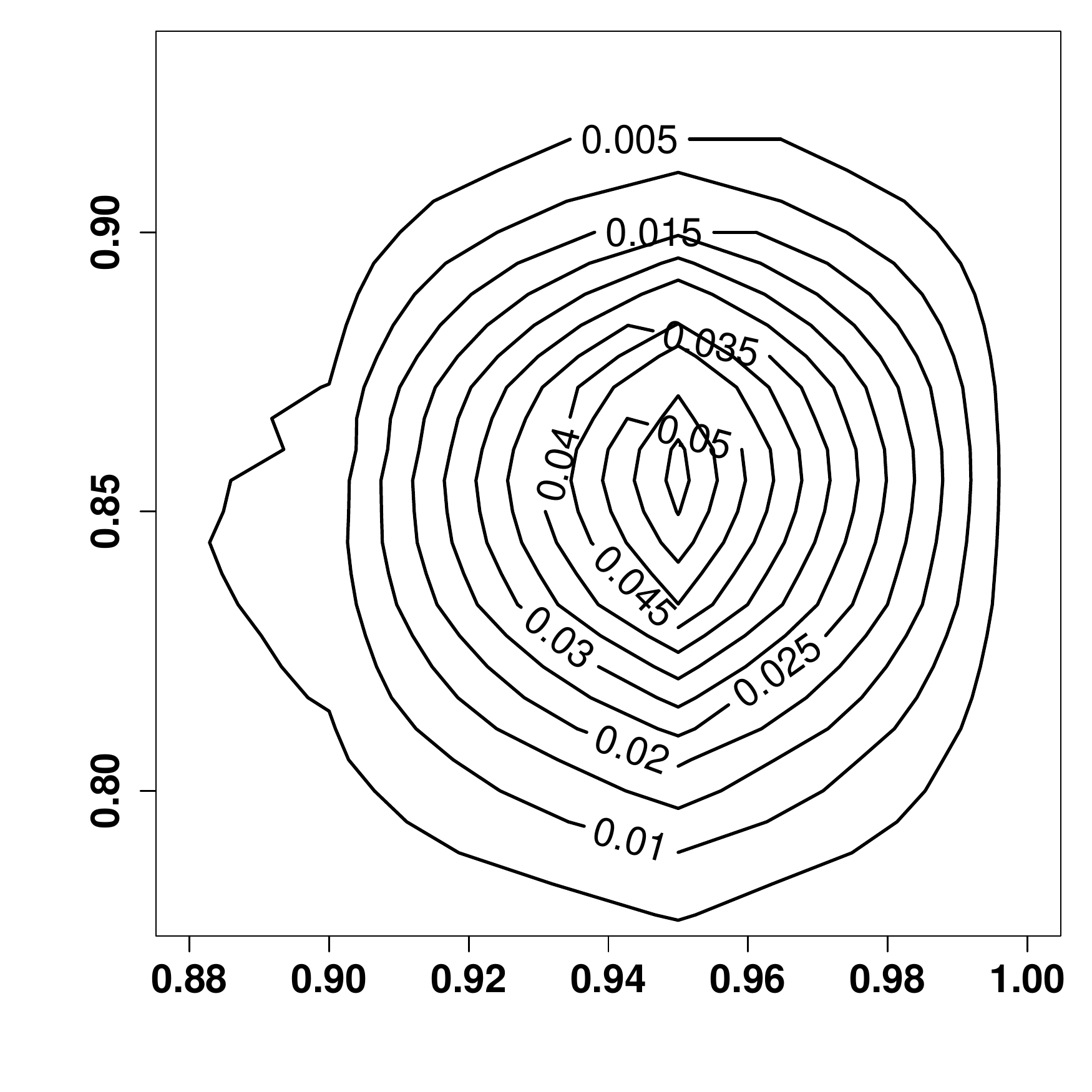}\label{fig:joint10}}
 \caption{Contour plots of proportions of compromised nodes with $(p_I,p_{II},q_I,q_{II})=(0.05,0.15,0.2,0.3)$. The $x$-axis presents the proportion of type I compromised nodes, and $y$-axis represents the proportion of type II compromised nodes.\label{fig:contour1}}
\end{figure*}

}
 \section{Conclusion and discussion}\label{conclusion}
Assessing the joint cyber risk of network systems is an important but challenging task. The challenge is primarily owing to three components: network topology, attack propagation, and heterogeneities of components. These integrated components result in interdependent cyber risks over the network. We propose  a novel backward elimination approach to efficiently computing the joint distribution of the number of compromised components over the network. {The developed backward elimination approach not only can provide explicit formulas for assessing the joint cyber risk of a small network but also
can be used to efficiently simulate the joint cyber risk of a large-scale network when the explicit computing is infeasible. }

We specifically introduce  a new concept of propagation depth $L$, which describes the power of risk propagation or the defensive power of a network. {It is rigorously shown that   when the propagation depth $L$ is larger, more nodes of all types over the network are compromised in the sense of the multivariate stochastic order. In particular, the number of compromised nodes is always stochastically larger when $L$ is larger for each type.} We further demonstrate  that when the compromise probabilities (direct or indirect) are larger, more nodes are compromised. The simulation study demonstrates that the number of compromised nodes increases significantly when $L$ increases from one to two. The correlation between the  proportions of compromised nodes was positive, as shown by the simulation study.

The results developed in this work can be used to score a cyber infrastructure with heterogeneous components, which can be further used for the purpose of risk management or cyber insurance. The current work can be extended in several directions. {For example, independence is assumed among all the propagation events in the current work. It would be interesting to study how the dependence among the propagation events would affect the risk propagation over the network. Our preliminary study shows that the dependence has a significant effect on the risk propagation. The other direction is to consider the propagation depth $L$ as a random variable because $L$ is indiscernible in certain practical situations. The theory of mixture models may be utilized to develop certain statistical inferences. Another possible study is the exploration of other propagation models (e.g., SIS or SIR) as the $L$-hop propagation model may not be suitable for certain networks/scenarios. Then, the joint cyber risk could be assessed based on the new propagation models, which would be different from the current work.}

 	\bibliographystyle{plain}
	\bibliography{hop,bayes}

\end{document}